\documentclass[sigconf,nonacm]{acmart}

\usepackage{amsmath,amsfonts,amssymb}
\usepackage{colortbl,xcolor}
\usepackage{graphicx}
\usepackage{hyphenat}
\usepackage[utf8]{inputenc}
\usepackage{listings}

\usepackage{mathtools}

\usepackage{mdframed}

\usepackage{microtype}
\usepackage{minted}
\usepackage{multirow}
\usepackage{sparklines}
\usepackage{varwidth}
\usepackage{csquotes}

\usepackage{float}
\usepackage{microtype}

\usepackage{xkeyval}

\AtBeginDocument{%
  \providecommand\BibTeX{{%
    \normalfont B\kern-0.5em{\scshape i\kern-0.25em b}\kern-0.8em\TeX}}}

\setcopyright{none}

\begin{document}
\raggedbottom  

\title[Learning\_to\_Program != “One-Size-Fits-All”]{Learning\_to\_Program != “One-Size-Fits-All”: Exploring Variations of Parsons Problems as Scaffolding}

\author{Carl Haynes-Magyar}
\orcid{0000-0002-9637-6285}
\affiliation{%
  \institution{University of Pittsburgh}
  \streetaddress{4200 Fifth Ave}
  \city{Pittsburgh}
  \state{Pennsylvania}
  \postcode{15260}
  \country{USA}
}
\email{cchmagyar@pitt.edu}




\renewcommand{\shortauthors}{Haynes-Magyar}

\begin{abstract}
Lowering the barriers to computer programming requires understanding how to scaffold learning. Parsons problems, which require learners to drag-and-drop blocks of code into the correct order and indentation, are proving to be beneficial for scaffolding learning how to write code from scratch. But little is known about the ability of other problem types to do so. This study explores learners' perceptions of a new programming environment called Codespec, which was developed to make computer programming more accessible and equitable by offering multiple means of engagement. Retrospective think-aloud interviews were conducted with nine programmers who were given the choice between Faded Parsons and Pseudocode Parsons problems as optional scaffolding toward solving write-code problems. The results showed that offering Faded and Pseudocode Parsons problems as optional scaffolds supported comprehension monitoring, strategy formation, and refinement of prior knowledge. Learners selectively used Faded Parsons problems for syntax/structure and Pseudocode Parsons problems for high-level reasoning. The costs noted included the time it takes to drag-and-drop the blocks and the confusion experienced when a solution diverges from a learners' mental model. Faded Parsons problems were also perceived as a desirable challenge. This study contributes to the field of computing education and human-computer interaction by extending the functionality of problem spaces that support Parsons problems and by providing empirical evidence of the effectiveness of using other problem types as scaffolding techniques.
\end{abstract}




\maketitle

\section{Introduction}
Learning to program can be difficult, but we can make it more achievable if we set realistic expectations and decenter cognitive norms for learners as a way of authoring positive trajectories of participation in computing education \cite{Du-Boulay1986-sl, Luxton-Reilly2016-bm}. In introductory computer programming courses, learners may experience high cognitive load because they lack the necessary schemata and plans to exploit while solving computer programming problems \cite{Robins2019-po}, or they may be negatively affected by poorly designed learning environments and difficult programming tasks that are introduced without the appropriate support \cite{Kelleher2005-ce}. For novice programmers taking computer science (CS) or CS-X courses (i.e., a non-CS major blended with the core computer science curriculum), the initial plunge into coding can be a significant rite of passage, one marred by cognitive overload and steep learning curves, but the rewards of learning how to compute for discovery, expression, or justice underscore the importance of innovative teaching practices and the acquisition of computing knowledge and skills \cite{Barr2023-pr}. The problem is that current introductory computer programming instruction and assessment may fail to scaffold the acquisition of these skills \cite{Xie2019-nr} and there is still more to learn about the relationship between skills (i.e., skill hierarchies) \cite{Pradhan-Newar2025-qh}. Computing education theorists posit these skills include: code reading and tracing, code writing, pattern comprehension, pattern application \cite{Lister2010-fi, Xie2019-nr}, reverse tracing, sequencing, and explaining code \cite{Pradhan-Newar2025-qh}.


\begin{figure}[H]
  \centering
  \includegraphics[width=230pt]{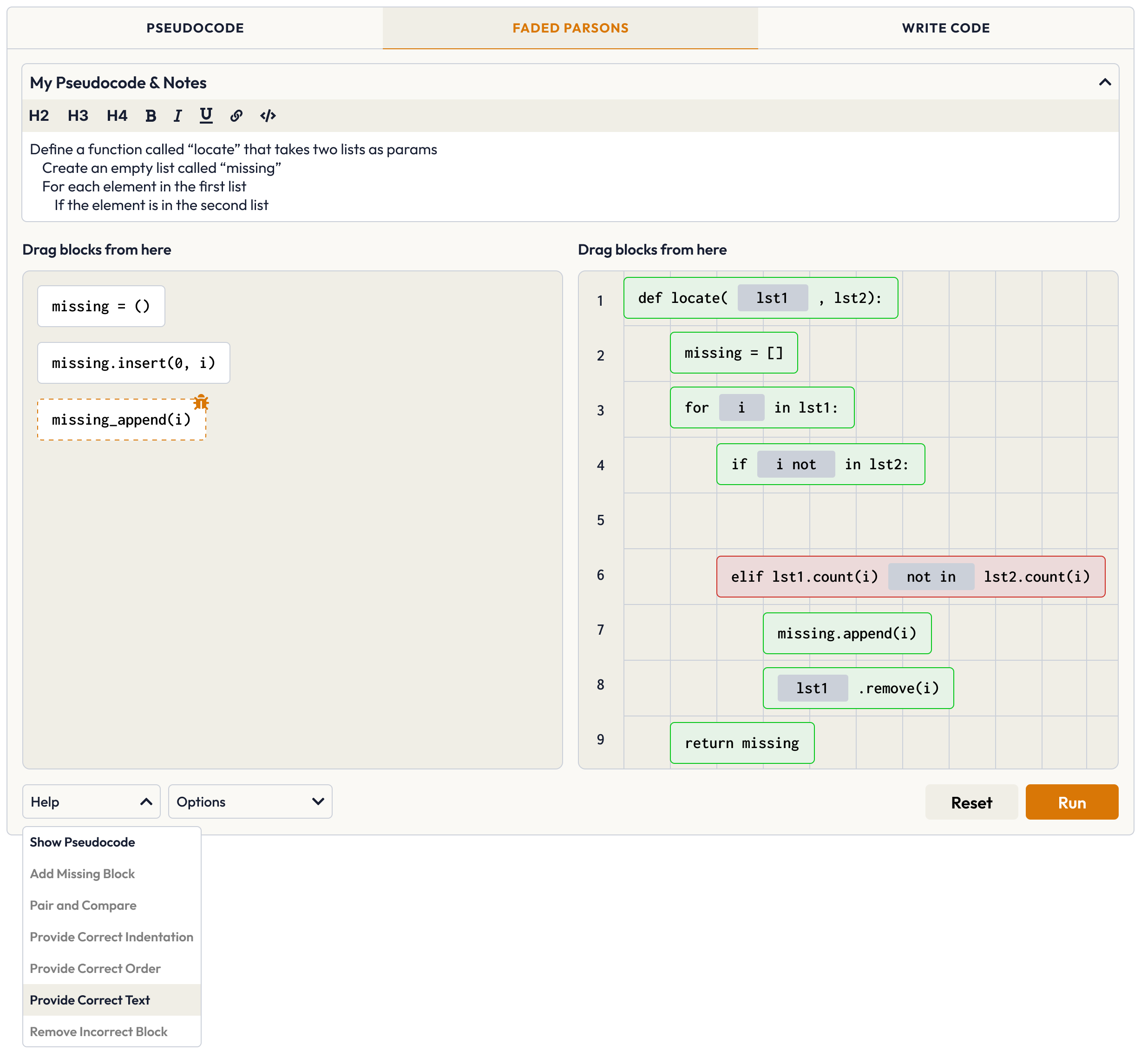}
  \caption{\href{https://www.codespec.org/}{Codespec's} problem space area featuring a Faded Parsons problem with a fix-code block denoted by the dashed orange/amber border and bug icon in the upper right-hand corner.}
  \Description{Type..}
   \label{fig:f1}
\end{figure}

It is within this challenging milieu that the pedagogical potential of Parsons problems emerges, offering a scaffolded learning environment that reduces cognitive load while underscoring conceptual understanding and the details of syntax \cite{Ericson2022-ab, Malmi2019-zu}. Traditionally, novice programmers have been instructed to practice programming by writing code from scratch\textemdash i.e., solving many small programs \cite{Allen2019-uj}. In contrast, Parsons problems require learners to construct a solution by rearranging and sometimes indenting blocks of code correctly, as shown in Figure \ref{fig:f1}.

Not all programmers are alike in their affinity for computer programming practice problems; some novice programmers reject code-tracing problems while others prefer it, and novice programmers with prior experience can have strong adverse reactions to Parsons problems \cite{Haynes2021-qd}, but they may find Faded Parsons \cite{Fromont2023-rn, Weinman2021-in} or Pseudocode Parsons problems (further described in Section~\ref{sec:relatedwork}) more suitable for scaffolding. Hence, there is a need to create tools and environments that reduce the barrier to programming by offering a wider range of problem types to challenge learners with and without prior programming experience \cite{Kelleher2005-ce}.

In response to this pedagogical exigency, this study explores learners' perceptions about choosing between different scaffolding techniques, such as Faded Parsons and Pseudocode Parsons problems. In this context, scaffolding means that when learners are asked to solve a write-code problem, they can choose to view and solve a Faded Parsons and or a Pseudocode Parsons problem to aid them in completing the write-code problem.

The research questions addressed in this study are:
\begin{enumerate}
    \item \label{rq:rq1} What do learners perceive are the advantages and challenges of using Faded Parsons and Pseudocode Parsons problems as optional scaffolding for write-code problems?
    \item \label{rq:rq2} What preliminary insights emerge from piloting the Scaffolding Learning Questionnaire with programmers using Codespec?
\end{enumerate}

This study extends the functionality of problem spaces that support Parsons problems through the development of Codespec\textemdash a new computer programming practice environment that offers multiple means of engagement (see in Section~\ref{sec:materials}). By examining how learners choose to engage with these scaffolds (i.e., different optional problem types), this study contributes to our understanding of their impact.
\section{Related Work}
\label{sec:relatedwork}
\subsection{Parsons Problems}
Parsons problems, originally called Parsons Programming Puzzles, were designed to: maximize engagement, constrain logic, permit common errors, model well written code, and provide immediate feedback \cite{Ericson2022-ab, Parsons2006-tw}. They can introduce learners to syntactic and semantic language constructs while exposing them to common programming patterns. Researchers have developed a variety of them since their inception. Parsons problems can vary by dimension, feedback, adaptation, use of distractor blocks, use of subgoal labels, use of fading, the number of lines, and use of optional blocks \cite{Du2020-ic, Ericson2019-ce, Ericson2022-ab, Morrison2016-qd, Oakeson2025-ev, Weinman2021-in}.

One-dimensional Parsons problems require learners to drag-and-drop blocks of code into the correct order only \cite{Denny2008-re}. Two-dimensional Parsons problems require learners to place blocks of code into the correct order \textit{and} indent them correctly \cite{Ihantola2011-np}. Research has shown that it is harder to solve two-dimensional as opposed to one-dimensional Parsons problems \cite{Denny2008-re} but the former has become the de facto template for introducing desirable difficulties \cite{Ericson2017-zo, Du2020-ic}.

Parsons problems can also be accompanied by different types of feedback, such as \textit{execution-based feedback} or \textit{line- or block-based feedback}. Execution-based feedback systems execute code and indicate correctness by returning expected and actual values and any error messages \cite{Helminen2013-su}. Line-based feedback systems highlight code blocks to indicate when they are in the wrong order or indentation, or when a code block needs to be rearranged somehow in order to create a working program. In a study by Helminen et al., the researchers found that ``it is not an either-or situation'' \cite[p. 59]{Helminen2013-su}. The responses to questions about whether or not learners thought line-based or execution-based feedback supported their learning were correlated. However, execution-based feedback caused learners to focus more on whether or not their code was executable, syntactically correct, and indented properly. Learners who received execution-based feedback requested feedback less frequently, but took more steps after the first feedback request to complete assignments. Importantly, the authors note that the evidence is not compelling for either type of feedback and that future research is needed \cite{Helminen2013-su}.

\begin{figure}[htbp]
  \centering
  \includegraphics[width=\columnwidth]{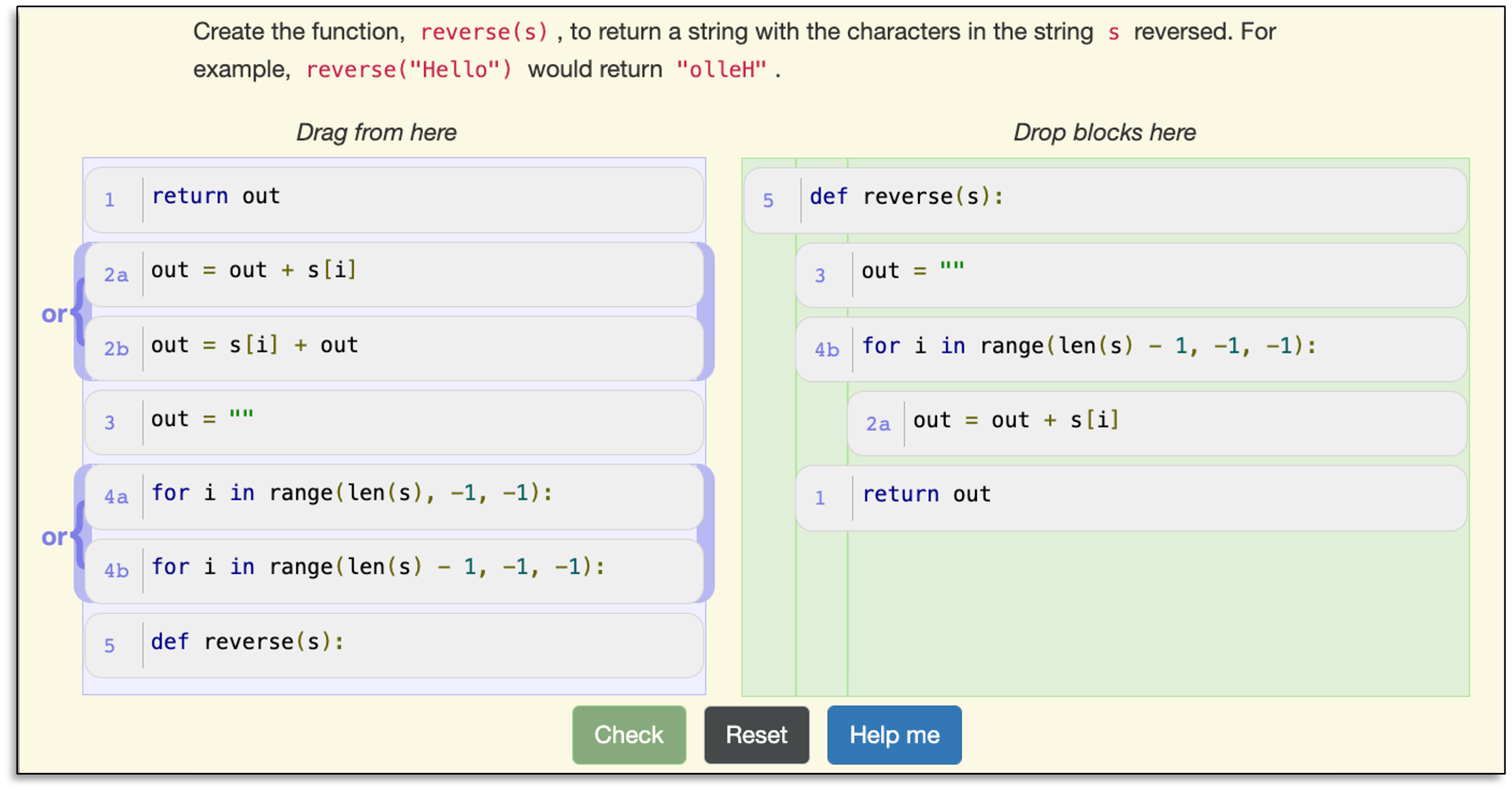}
  \caption{This is an example of a two-dimensional Parsons problem with paired distractor blocks from \href{https://runestone.academy/user/login?_next=/}{Runestone Academy}.}
  \Description{Type...}
  \label{fig:paired_distractor}
\end{figure}

Furthermore, Parsons problems can be adaptive \cite{Ericson2018-bp}. First, \emph{intra}-problem adaptation modifies the difficulty of a Parsons problem by removing a distractor block or combining blocks into one when a learner clicks on a ``Help'' button. And, second, \emph{inter}-problem adaptation modifies the difficulty of the next Parsons problem based on a learner's prior Parsons problem performance. If a learner struggles, a subsequent problem is made easier by removing some or all of its distractor blocks and pairing them with the correct code. Conversely, if a learner solves a Parsons problem in just one attempt, the subsequent one is made harder by adding and jumbling distractor blocks. Distractors are code blocks that are not part of the correct solution and can be paired or unpaired with its correct counterpart (see Figure \ref{fig:paired_distractor}) \cite{Smith2023-em}. Studies have shown that solving adaptive Parsons problems is a more efficient and equally effective form of practice as solving traditional write-code problems \cite{Ericson2018-bp}, but not when the Parsons problem solution is unusual \cite{Haynes2021-qd, Haynes-Magyar2022-tr}. Adaptive Parsons problems have also proven to be an effective active learning technique during lecture that learners have found more helpful for their learning than write-code exercises \cite{Ericson2022-hp}. Finally, research on the use of distractor blocks has shown that they are ill-suited for summative assessments as they substantially lengthen time-on-task yet yield little to no improvement in problem discrimination \cite{Smith2023-ep, Smith2023-pv}.

\begin{figure}[H]
  \centering
  \includegraphics[width=230pt]{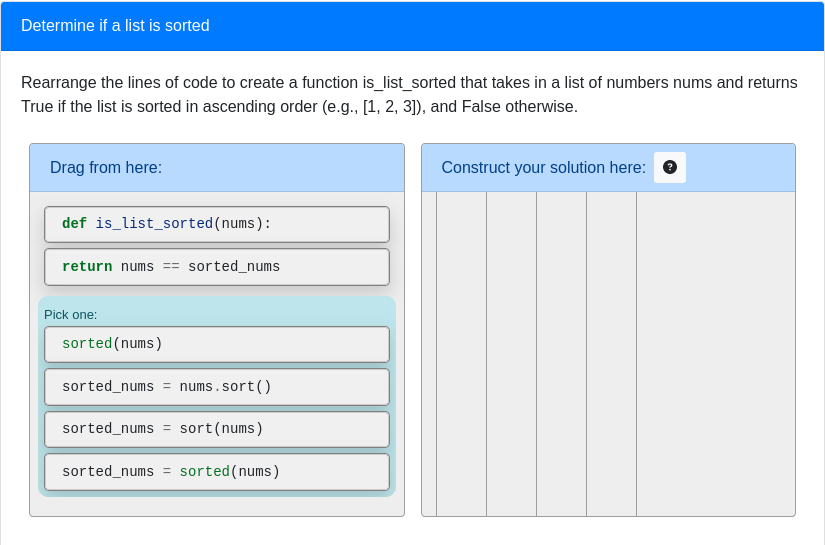}
  \caption{This is an example of a two-dimensional Parsons problem with unpaired distractor blocks from \href{https://us.prairielearn.com/}{PraireLearn} that was used in \cite{Smith2024-cz}.}
  \Description{Type..}
   \label{fig:smith2024-figure1}
\end{figure}

Subsequent work, in a proctored computer lab, similarly found that adding distractors\textemdash whether visually pairing them with their correct code block or jumbling them\textemdash adds to time-on-task and reduces students' performance on quizzes and exams; pairing them does somewhat mitigate the time cost, but not enough to resolve concerns \cite{Smith2023-em}. By contrast, exposure to Parsons problems with distractors outside of a controlled lab setting resulted in a large margin of near and far learning transfer over learning conditions without distractors, and think-aloud observations suggested that paired distractor blocks support learners in attending to the details of code (see Figure \ref{fig:smith2024-figure1}) \cite{Smith2024-cz}. Furthermore, there is evidence that neurodiverse learners (e.g., with attention-deficit/hyperactivity disorder or Tourette syndrome) may benefit more from \textit{paired} than from \textit{jumbled} distractor blocks \cite{Haynes-Magyar2024-au}. Researchers have also explored the quality of distractor blocks generated by human beings versus by large language models (LLMs) and found that GPT-4o performed best, reproducing roughly half of the human-written functional distractors \cite{Hassany2025-kn}. Quality was based on whether or not distractor blocks were more effective\textemdash meaning ``those that are plausible (often picked by students) and potentially based on common misconceptions'' \cite[p. 484]{Hassany2025-kn} Also, all of the LLMs used in the study generated novel distractor blocks that were different from pre-existing human-written distractor blocks \cite{Hassany2025-kn}.

\begin{figure}[h!]
  \centering
  \includegraphics[width=230pt]{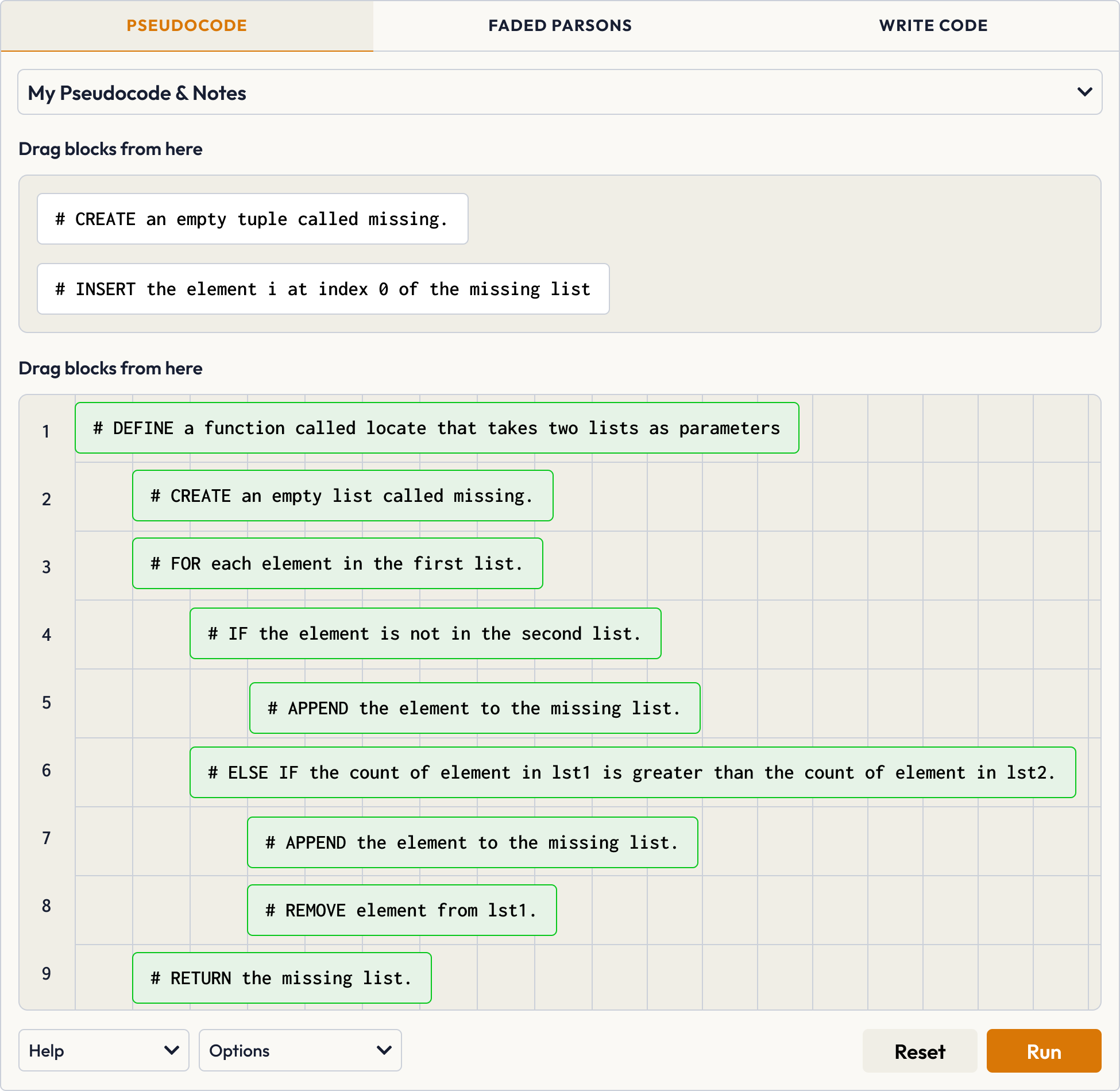}
  \caption{\href{https://www.codespec.org/}{Codespec's} problem space area featuring a Pseudocode Parsons problem.}
  \Description{Type..}
   \label{fig:f2}
\end{figure}

Subgoal learning is a strategy that can help students solve Parsons problems by breaking down problem-solving procedures into functional parts, or subgoals of a problem solution \cite{Catrambone1998-ka}. A subgoal can represent ``a meaningful conceptual piece of an overall solution procedure'' \cite[p. 357]{Catrambone1998-ka}. It is akin to pseudocode, which programmers use to convey the specific steps of an algorithm in plain English so that anyone with basic programming knowledge can understand. Programmers refer to pseudocode as ``informal textual representations of a program or algorithm'' \cite[p. 227]{Bellamy1994-wo}. Parsons problems can be accompanied by subgoal labels that are visually paired with code blocks \cite{Morrison2016-qd}. For example, the first block on the right in Figure \ref{fig:f1} would have a subgoal label in the form of a comment that reads ``\# DEFINE a function called locate that takes two lists'' along with the code \textbf{\texttt{def locate (lst1, lst2):}}. Solving one-dimensional Parsons problems with subgoal labels has led to significantly more learning gains for understanding how to write, for example, a \textbf{\texttt{while}} loop than solving Parsons problems with self-generated or no subgoal labels \cite{Morrison2016-qd}. This study introduces Pseudocode Parsons problems (see Figure \ref{fig:f2}) which are similar in purpose to subgoal labels \cite{Catrambone1998-ka, Morrison2016-qd}, Explain in Plain English (EiPE) questions \cite{Murphy2012-ey, Smith2024-tr}, Design-level Parsons Problems \cite{Garcia2018-at, Garcia2021-jt}, and Prompt problems \cite{Denny2024-kx}. Each of these problem types aim to help learners read, comprehend, and explain each line of code and how it functions.


Faded Parsons problems, a recent variation, require learners to fill in the blanks within code blocks \cite{Weinman2021-in}. The goal is to expand on Parsons problems' ability to scaffold the acquisition and application of programming patterns and to provide a challenge to learners with more prior programming experience. This also helps keep learners in Vygotsky's Zone of Proximal Development (ZPD), which is the difference between what learners can do independently and what learners can do with assistance \cite{Vygotsky1978-lj}. Weinman, Fox, and Hearst found Faded Parsons problems efficacious in (1) teaching learners to identify and apply programming patterns when relevant and (2) facilitating learning transfer; they also found learners preferred Faded Parsons over code-writing questions because of the perceived difficulty of the latter and the fact that Faded Parsons problems ``allow[ed] them to think like the instructor'' \cite[p. 10]{Weinman2021-in}.

\begin{figure}[htbp]
  \centering
  \includegraphics[width=\columnwidth]{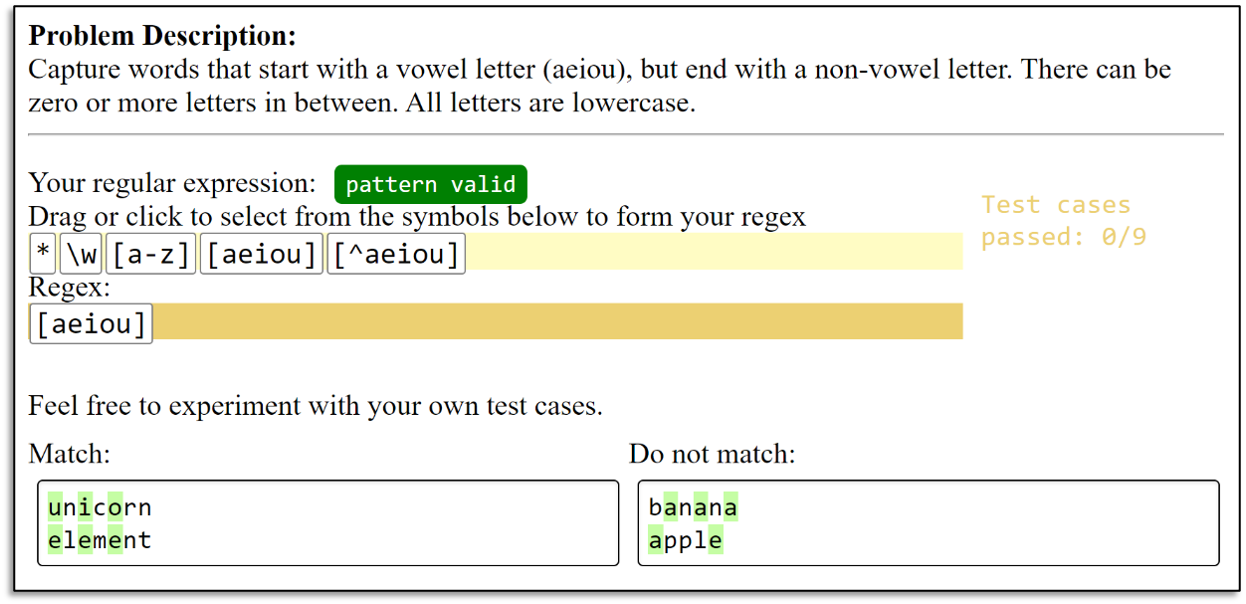}
  \caption{This is an example of a micro Parsons problem from \cite{Wu2023-ui}.}
  \Description{Type...}
  \label{fig:wu2023-figure1}
\end{figure}

Parsons problems can also vary by the number of lines a learner has to interact with and the number of solutions that are possible. Micro Parsons problems focus on a single line of code and require learners to assemble code fragments into the correct order (see Figure \ref{fig:wu2023-figure1}) \cite{Wu2023-ui}. Evidence shows learners engage in solving micro Parsons problems, with execution-based feedback, significantly more than code-writing problems and that there are learning gains \cite{Wu2023-ui}. Learners also vary in their preferences for micro Parsons problems with block-based feedback and execution-based feedback over traditional code-writing problems; research has shown that learners who practice by solving the former reap significantly higher learning gains than those who practice by solving the latter \cite{Wu2024-ni}. Furthermore, Wu and Smith IV used micro Parsons problems as exam items and drew on measurement theory to compare their psychometrics to traditional single-line code-writing problems \cite{Wu2024-kx}. They observed parity in difficulty but slightly lower item discrimination for micro Parsons, and learners had mixed preferences for the different problem types as exam questions\textemdash their preferences were based on the potential to earn more points \cite{Wu2024-kx}. Also, there was evidence of the expertise reversal effect. Some learners considered solving micro Parsons problems harder than solving code-writing problems \cite{Wu2024-kx}. And last, but not least, researchers are exploring how we can extend the functionality of computer programming practice problem spaces that support Parsons problems by creating optional blocks. These types of blocks ``[afford] students a method to engage with questions that have several valid solutions composed of different answer components'' \cite[p. 1]{Oakeson2025-ev}. This study also introduces two new features to the functionality of problem spaces that support Parsons problems. First, Codespec supports the options to `Add Blocks' or `Copy Blocks.' a learner can add a block in order to create their own solution\textemdash building on the blocks that are already available\textemdash or copy blocks of code that they want to use in their solution for a different problem type. The `Copy Blocks' feature supports learners in beginning to solve a Parsons problem and transferring that solution over to the write-code problem.

Various programming practice websites and open-source tools support instructors and learners in using Parsons problems, including \href{https://www.codio.com/}{Codio}, \href{https://www.codespec.org/}{Codespec}, \href{https://epplets.org/}{Epplets}, \href{https://js-parsons.github.io/}{js-parsons}, the \href{https://doi.org/10.1145/3279720.3279726}{PCEX} system, \href{https://us.prairielearn.com/}{PraireLearn}, \href{https://runestone.academy/user/login?_next=/}{Runestone Academy}, and \href{https://github.com/HiruNya/Parsons-Tool}{UPP (the Unnamed Parsons Problem Tool)}.

Using Parsons problems to scaffold learning was the result of Haynes and Ericson finding these problems can aid college learners in solving equivalent write-code problems \cite{Haynes2021-qd, Haynes-Magyar2022-tr} and early related research on blocks-based and text-based programming conducted by Weintrop \cite{Lin2021-dx} The first study to use Parsons problems to scaffold solving write-code problems provided evidence that they are beneficial for scaffolding learning how to write code from scratch \cite{Hou2022-xa}. The benefits of using Parsons problems as scaffolding included minimizing difficulty and time-on-task, exposure to new programming strategies, updates to current programming knowledge, and deeper awareness about one's learning process. Challenges include struggling to resolve Parsons problems because of their solution or distractor blocks, there not being a broad range of scaffolding options, and help-avoidance. Notably, some systems only offer help after three attempts, which may lead to a trial-and-error problem-solving strategy. A recent study using Parsons problems as scaffolding for learning difficult programming concepts resulted in significantly higher practice performance and problem-solving efficiency for learners with low self-efficacy in computer science \cite{Hou2023-dg}. However, the latter study also provides evidence that learners with high self-efficacy prefer to write code from scratch and use Parsons scaffolding sparingly. Hence, not all programmers are alike in their scaffolding needs for computer programming practice.

The inclusion of other problem types, such as Faded and Pseudocode Parsons problems for scaffolding write-code tasks, has not been explored. Recall that Faded Parsons problems require learners to complete blocks of code with partially blank lines and order them into a valid program \cite{Weinman2021-in}. These problems are significantly more effective for teaching pattern comprehension and pattern application over code-writing and code-tracing exercises and comparably improve students' ability to write code from scratch \cite{Weinman2021-in}. Instructors can also modify Faded Parsons problems to easily provide each learner with programming practice problems that are appropriate for the learner's level of expertise \cite{Fromont2023-rn}. Furthermore, the creation/inclusion of Pseudocode Parsons problems (see Figure \ref{fig:f2}) is organic given that their instructional goal of making program structure and intent explicit at a readable level is similar to that of subgoal labels, Explain-in-Plain-English (EiPE) questions, Design-level Parsons problems, and Prompt problems. Bellamy notes that the term `pseudo-code' usually refers ``to informal textual representations of a program or algorithm'' \cite[p. 277]{Bellamy1994-wo}. Pseudocode Parsons problems present line-by-line algorithmic intent in plain language and require learners to align that intent with code structure, thereby supporting decomposition (as with subgoal labels) \cite{Catrambone1998-ka, Morrison2016-qd}, articulation of codes purpose (as with EiPE) \cite{Murphy2012-ey, Smith2024-tr}, design-level reasoning (as in Design-level Parsons) \cite{Garcia2018-at, Garcia2021-jt}, and guided formulation of solutions (as in prompt-driven tasks) \cite{Denny2024-kx}.

\subsection{Cognitive Load}
Cognitive load is defined as the tax on working memory that has an impact on the processing of new information during a learning task \cite{Sweller2019-ag}. Cognitive Load Theory distinguishes between intrinsic load (arising from the inherent difficulty of the learning task), extraneous load (stemming from the instructional design and presentation of the learning task), and germane load (the mental effort required during the learning task), though the latter is up for debate \cite{Paas2010-xi}. Learning tasks can be designed to reduce cognitive load by reducing the complexity of the task, improving the clarity of instructions, and by the designing accessible interfaces\textemdash all of which are mediated by the learner’s prior knowledge \cite{van-Merrienboer2005-eb}. Furthermore, complexity of the learning task can be determined by its element interactivity \cite{Chen2023-gj, Sweller2010-uq}. Element interactivity refers to the complexity of learning new concepts that rely on having amassed prior knowledge of other concepts \cite{Chen2015-wr}; learning new information can require the simultaneous consideration of a vast number of elements that interact or that do not interact \cite{Sweller2025-bo}. Programming can result in high cognitive load for programmers when they lack the schemata necessary to free up working memory \cite{Robins2019-po}. 

Regarding task difficulty and cognitive load, researchers posit ``A difficult task does not necessarily lead to greater mental effort.  Unless the individual is interested in the task and wants to do it well, task difficulty would lead the individual to give it up instead of trying to invest further effort.  Thus, conceptually, a task with high cognitive load should be a task that is perceived to be difficult and cannot be done well even though the individual likes the task and invests a high level of effort in it'' \cite{Yeung2000-bs}.

Evidence of cognitive load has been gathered indirectly, directly, subjectively, and through dual-task performance measures \cite{Klepsch2017-bx, Robins2019-iv}.  Scientists have developed scales that are both unitary (combining categories) and differential (subscales for each category) \cite{Klepsch2020-er, Zheng2017-lr}. The most valid and reliable measure is the Paas scale \cite{Paas1992-mk}. Computing education researchers have developed a differential scale from the Cognitive Load Component Survey (CLCS) \cite{Morrison2014-zv}, but it has provided mixed results \cite{Harms2016-bx, Morrison2014-zv, Zavgorodniaia2020-uw}.  Other validated measures of cognitive load include: the NASA Task Load Index (NASA-TLX) \cite{Hart1988-tc}, the SOS scale \cite{Eysink2009-kl, Swaak2001-rf}, and the differentiated cognitive load measure developed by Klepsch, Schmitz, \& Seufert \cite{Klepsch2017-bx}.

\subsection{Scaffolding}
\label{sec:scaffolding}
Scaffolding is vital in educational contexts, especially in computing education where the cognitive load of mastering new programming skills and knowledge can be overwhelming \cite{Kelleher2005-ce}. Scaffolding helps learners participate in activities that share elements of expert practices without requiring knowledge and skills that learners do not yet possess \cite{Reiser2014-ql}. Wood et al. first defined scaffolding as a ``process that enables a child or novice to solve a problem, carry out a task or achieve a goal which would be beyond [their] unassisted efforts'' \cite[p. 90]{Wood1976-sr}.

Researchers have proposed several ways to categorize scaffolding. First, scaffolding can be organized into two categories according to how it is delivered: tacit or explicit \cite{Hadwin2001-oe}, hard or soft \cite{Saye2002-lm}, or fixed or adaptive \cite{Azevedo2005-ha}. Tacit scaffolds are tools designed to subtly draw learners' attention to their study process through implicit cues rather than through explicit direction or instruction. Hard scaffolds are fixed, pre-designed supports that instructors prepare in advance to address the common challenges learners typically encounter with a given task. In contrast, soft scaffolds are flexible and context-dependent\textemdash they are adaptive. This approach requires instructors to continuously assess what learners understand and offer responsive, timely assistance that adapts to their needs. Finally, fixed scaffolds, although easy to embed, do not adjust to learners' needs and adaptive scaffolding can be offered by a human \cite{Chi2001-ix, Chi2004-vf} or computerized tutor \cite{Aleven2002-im} to meet individual learning needs.

Second, scaffolding can also be divided into three categories: supportive, reflective, or intrinsic \cite{Metcalf2012-ey}, or peer, teacher, or technology \cite{Kim2011-ms}. And, third, researchers have proposed four types of scaffolding: conceptual, metacognitive, procedural, and strategic \cite{Hannafin2013-wa}.

The emphasis of this study was on cognitive scaffolding, which focuses on how to provide gradual assistance to the learner to solve problems, and metacognitive scaffolding, which cues the learner to reflect on their own learning \cite[cf.,][]{Huang2025-sj}. Key components of cognitive and metacognitive scaffolding supported strongly in the literature by Reiser and Tabak \cite[p. 48-50 \S \href{https://doi.org/10.1017/CBO9781139519526.005}{3.3 How Can Scaffolding Transform Learning Tasks?}]{Reiser2014-ql} are described below and were used to guide the creation of a new questionnaire. Scaffolding can:

\begin{enumerate}
    \item help simplify the elements and element interactivity of a task so that it is accomplishable,
    \item provide strategic support during the learning process so that learners can experience what it’s like to work and learn in an authentic environment,
    \item offset frustration and help with retention,
    \item help learners identify critical features of a task that they may overlook,
    \item prompt learners to reflect on their learning and articulate their ideas more effectively,
    \item and facilitate practical learning in a real-world context.
\end{enumerate}

Notably, a gap exists in the instruments for evaluating the effectiveness of learning technologies to scaffold learning, especially in the context of introductory computer programming courses \cite{Zavaleta-Bernuy2020-ra}. This paper aims to bridge this gap by piloting the Scaffolding Learning Questionnaire. Effective scaffolding can significantly enhance the learning experience and, if measured alongside learning gains, may operationalize how we track of Vygotsky's Zone of Proximal Development (ZPD)\textemdash the difference between what learners can do autonomously and what learners can do with support \cite{Vygotsky1978-lj}.

\section{Methodology}
\begin{table*}[ht!]
\caption{Participant Demographics}
\label{tab:demographics}
\centering
\begin{tabular}{l l>{\raggedright\arraybackslash} p{0.1\textwidth} >{\raggedright\arraybackslash}p{0.15\textwidth} >{\raggedright\arraybackslash}p{0.25\textwidth} >{\raggedright\arraybackslash}p{0.25\textwidth}}
\hline
\textbf{ID}  &\textbf{Age}  &\textbf{Gender}    &\textbf{Ethnicity}   &\textbf{University / Major}   &\textbf{Programming Experience}\\ \hline
\rowcolor[HTML]{EFEFEF}
Learner\_020     &20     &Male   &Asian                          &Stanford University / Computer Science                                            &1 yr. 2 mos.   \newline   Python, C++\\
Learner\_021$\ast$     &21     &De-Identified  &Hispanic, Latino or Spanish    &De-Identified / Computer Science                                            &3 yrs.         \newline   JavaScript, HTML/CSS, Python, C++\\
\rowcolor[HTML]{EFEFEF}
Learner\_022     &18     &Male   &Asian                          &University of California, Berkeley / Electrical Engineering \& Computer Science   &3 yrs.         \newline   Python, C\#\\
Learner\_023     &23     &Male   &Asian                          &University of California, Berkeley / Electrical Engineering \& Computer Science   &3 yrs.         \newline   JavaScript, HTML/CSS, SQL, Python, Java, C++\\
\rowcolor[HTML]{EFEFEF}
Learner\_024$\ast$     &37     &De-Identified   &De-Identified         &De-Identified / Applied Data Science                                     &13 yrs.        \newline   JavaScript, HTML/CSS, SQL, Python, Java, C\#, R\\
Learner\_025$\ast$     &23     &Female &Asian, White                   &De-Identified                                                                     &7 yrs.         \newline   HTML/CSS, SQL, Python, Java\\
\rowcolor[HTML]{EFEFEF}
Learner\_026     &18     &Male   &Asian                          &Stanford University / Computer Science                                            &6 yrs. 1 mon.  \newline   JavaScript, HTML/CSS, Python, C++, C\\
Learner\_027     &21     &Male   &Asian                          &Carnegie Mellon University / Educational Technology \& Applied Learning Sciences  &3 yrs.         \newline   JavaScript, HTML/CSS, SQL, Python, C++\\ 
\rowcolor[HTML]{EFEFEF}
Learner\_028     &19     &Male   &Asian                          &University of California, Berkeley / Electrical Engineering \& Computer Science   &3 yrs.         \newline   HTML/CSS, SQL, Python, Java\\ 
\bottomrule
\end{tabular}

    \vspace{0.5ex}

    \begin{minipage}{\linewidth}
        \footnotesize
        \textit{Note:} To protect participant privacy, select demographic information was de-identified for participants marked with an asterisk $\ast$. De-identification decisions were guided by k-anonymity principles, which assess re-identification risk based on unique combinations of quasi-identifiers \cite{Morehouse2025-nm}.
    \end{minipage}
\end{table*}
The study received institutional review board (IRB) approval to recruit participants via email and flyers from post-secondary institutions and listservs in the United States. Inclusion criteria were for novice programmers with little experience writing in Python as determined by a prior programming experience survey, and for participants for whom codespec.org could currently support their interaction with the learning environment.

\subsection{System Design}
Few computer programming environments offer learners the option to solve the same problem across different problem types. Codespec, used in this study, is a computer programming environment that supports learners in solving a programming problem as a Pseudocode Parsons problem, a Parsons problem, a Faded Parsons problem, a fix-code problem, or a write-code problem \cite{Haynes-Magyar2022-oj, Bunde2023-rw}.

It also features learner-initiated help-seeking options. When a learner clicks on the `Help' button, they can choose to: `Show Pseudocode,' `Pair and Compare' a correct and distractor block to remove the latter, `Remove Incorrect Blocks' that are not paired with a distractor block, `Add a Missing Block,' `Provide the Correct Order,' or `Provide the Correct Indentation.' It also features `Options' to `Add Blocks' or `Copy Blocks' and a new type of block called fix-code blocks, which have bugs that need to be corrected. The `Add Blocks' option ensures learners can build on or create their own solution and the `Copy Blocks' option ensures learners can copy code that they want to use in their solution for a write-code or fix-code problem. 

Codespec also supports instructors in creating different solutions for each of the problem types and in other programming languages, such as JavaScript, to challenge learners and expand on instructors' ability to personalize and adapt the learning environment to each individual learners' needs.

\subsection{Materials}
\label{sec:materials}

The problems covered a range of programming fundamentals (see Table \ref{tab:problem_set}): list manipulation and searching, conditional logic and edge cases, looping and condition checking, string manipulation, algorithmic thinking, data structure usage, function design and modularity, boolean logic, return values and output formatting, and range and index operations. Following each problem, participants were asked to complete the Paas scale \cite{Paas1992-mk}. This question asks participants to rate how much mental effort they invested in solving a particular problem using a nine-point Likert scale: 1 = \textit{''very, very low mental effort,''} 2 = \textit{``very low mental effort,}'' 3 = \textit{``low mental effort,''} 4 = \textit{``rather low mental effort,''} 5 = \textit{``neither low nor high mental effort,''} 6 = \textit{``rather high mental effort,''} 7 = \textit{``high mental effort,''} 8 = \textit{``very high mental effort,''} 9 = \textit{``very, very high mental effort.''} Upon solving all of the problems, participants were also asked to respond the Scaffolding Learning Questionnaire using a five-point Likert scale of \textit{``strongly disagree''} to \textit{``strongly agree.''} The questionnaire had eight items that the researcher developed from the principles outlined in Section \ref{sec:scaffolding}.

\begin{table}[ht]
    \centering
    \caption{Problem Set and Categories Covered}
    \label{tab:problem_set}
    \begin{tabular}{ll}
        \toprule
        Category & Problems \\
        \midrule
        List Manipulation           & One, Five, Eleven \\
        String Manipulation         & Eight, Nine \\
        Counting/Iteration          & Three, Four \\
        Middle Element Extraction   & Two, Ten \\
        \bottomrule
    \end{tabular}
\end{table}

\subsection{Procedure}
Participants were asked to: (1) answer questions about their demographics, prior programming experience \cite{Holden2003-oi, Siegmund2014-zj}, and learning preferences, and (2) solve eleven computer programming problems. The programmers were given ten minutes to solve each problem. This heuristic estimation was based on prior research that found it takes five minutes on average to solve either a Parsons or write-code problem \cite{Haynes-Magyar2022-tr}. The Paas scale was administered after each problem \cite{Paas1992-mk}. The scale consists of one question that asks respondents to rate how much mental effort they invested in solving a problem; it uses a 9-point Likert scale from ``very, very low mental effort'' to ``very, very high mental effort.'' Upon completing all of the programming problems, I conducted a retrospective think-aloud interview during which the participants had access to the problems for recollection.

\subsection{Participants}
Eight participants identified as male, one identified as female; eight identified as Asian, and one as Hispanic (see Table \ref{tab:demographics}). The ages ranged from 18 to 37 years old ($M = 22$ years old, $SD = 5.85$). One participant self-identified as having a disability—an attention deficit hyperactivity disorder (ADHD). Five participants were Computer Science (CS) majors, one was an Electrical Engineering and Computer Science major, one was an Applied Data Science (DS) major, one was a Symbolic Systems major, and one was in an Educational Technology and Applied Learning Sciences major. The participants had experience with a range of programming languages (i.e., JavaScript, HTML/CSS, SQL, Java, C\#, C++, C, R) that spanned between 1 to 13 years ($M = 56.33$ months, $SD = 43.06$, $Mdn = 36$ months)\textemdash all of the participants were fairly new to programming in Python. Participants were also asked what computer programming concepts they struggle with, and the responses included: algorithms, databases (indexing, lock), coroutine, design (DDD, clean architecture), list comprehension, unfamiliar with many external libraries, regex, graphs, dynamic programming, bit manipulation, user-defined functions, floating-point numbers, functional programming, object-oriented programming, responsive web development, decorators, and remembering to use exception handling.

All the participants were asked the following question about their preferences: ``When you are learning something new, do you prefer to (1) have someone show you how to do it, (2) have someone tell you how to do it, (3) figure it out yourself, or (4) other, please specify. Five participants selected ``figure it out yourself;'' two selected ``have someone tell you how to do it,'' one selected ``have someone show you how to do it,'' and one selected option four and wrote, ``Try it once myself and then be given hints towards the solution.''

\subsection{Analysis}
Audio and video were recorded via Zoom and transcribed using Whisper, a pre-trained model for automatic speech recognition (ASR) and speech translation. Qualitative analysis was performed using ATLAS.ti. The data was analyzed using a deductive approach. Codes were derived and refined based on the headings from a recent paper that explored students' perceived benefits and challenges of using Parsons problems to scaffold write-code problems \cite[p. 21]{Hou2022-xa}. I trained an undergraduate research assistant to independently code 100\% of the transcripts and identified examples independently until we reached 100\% agreement.
\section{Results}

\subsection{Retrospective Think-Aloud Interviews}
\subsubsection{Advantages of using Faded and Pseudocode Parsons problems as scaffolding} To better understand participants' perceptions of the advantages and challenges of using Faded and Pseudocode problems to scaffold write-code problems, a retrospective think-aloud interview was conducted with each participant after they solved all of the problems. First, the perceived advantages are presented.

Comprehension monitoring is a metacognitive strategy that involves an individual's ongoing awareness and assessment of their understanding \cite{Loksa2020-cy}. Two participants reported comparing their solution to both the Faded and Pseudocode Parsons problems after completion. For example, Learner\_025 said, ``...one benefit of using both [problem types] was that I could write my own solution first and then go back to the Faded Parsons and the Pseudocode Parsons and check if the syntax there or the overall general logic matched with what I'd already written...it was a confirmation that I was on the right track....as a person who has programmed quite a bit before it helps me to actually write out my idea first and then compare it with existing approaches to see if it's efficient or if there's a shorter way or an easier way to do it as well.''

Challenging Problem Type: Two participants preferred the Faded Parsons instead of Pseudocode Parsons problems because the fill-in-the-blanks made them more challenging. For instance, Learner\_028 said, ``I think the Faded Parsons problem is interesting because it doesn't give you all the answers and it helps you see what variables are necessary and how you could structure your code to arrive at a solution. I think for a first-time programmer that'd be helpful, especially when you're not too sure about how you would write out your solution given the instructions\textemdash it's sometimes better to start with a Faded Parsons problem solution and go back to the write-code problem to figure out how you would write it.''

Other advantages were consistent with findings described by \cite{Hou2022-xa}. Participants reported that advantages to using Faded and Pseudocode Parsons problems were: (1) to reduce difficulty and completion time, (2) to learn problem-solving strategies, (3) to refine and extend existing programming knowledge, and (4) that they prompted them to think more deeply. For example, Learner\_026 said about learning problem-solving strategies, ``I used to teach high schoolers computer programming and the part they struggled with was not the syntax, but mostly the pseudocode itself\textemdash the problem-solving process itself. They don't know how to piece these things together. Syntax is not what they struggle with because you can teach syntax. I think a lot of time I taught too much about syntax and not enough about problem-solving. I would keep Pseudocode Parsons for beginners instead of Faded Parsons.'' This was also understood as a disadvantage of using Faded Parsons problems\textemdash they give away the syntax. Another participant, Learner\_024, said about refining and extending programming knowledge that, ``I used the Faded Parsons problems because they gave me hints about coding, it described things in Python code, so I was able to directly understand the code I wrote and the Faded Parsons problem's code.'' About the advantages of Faded over Pseudocode Parsons problems, Learner\_026 said, ``The main benefit was that if I didn't know a specific function....it was really helpful to have the Faded Parsons problem rather than pseudocode because it's more visual.''

\subsubsection{Challenges of using Faded Parsons and Pseudocode Parsons problems as scaffolding} While Parsons problems have significantly aided learners in developing coding skills, several participants perceived challenges. Two participants reported that dragging-and-dropping Parsons problems blocks increased completion time. Learner\_022 said, ``Usually, when I try to debug, I use the print statement or the debug tools and the IDE....I'm not familiar with debugging or figuring things out with Faded and Pseudocode Parsons problems. I felt that they were a bit confusing and a bit complex to use because you needed to drag-and-drop the blocks....there was no benefit to using them for me.'' And Learner\_021 said, ``I didn't find much benefit in using the blocks themselves because it would take longer for me to drag-and-drop them.''

Another participant said, about having difficulty understanding the Parsons solution, ``supposing I had some idea already in my head, but it wasn't super clear, and I wanted to look over the options for the blocks in the pseudocode problem. If some of the blocks did not correspond to the idea that I had in my head for the program, it would get a bit confusing. Especially since the blocks are out of order and you have to spend more mental effort to rearrange those blocks....Overall, using the Pseudocode Parsons problem was a lot more intuitive for me than using the Faded Parsons problem [because] it gave me a better logical overview of the program [when it match what I was thinking]'' (Learner\_025).

\subsection{Cognitive Load Ratings}
\begin{figure*}[htbp]
  \centering
  \includegraphics[width=500pt]{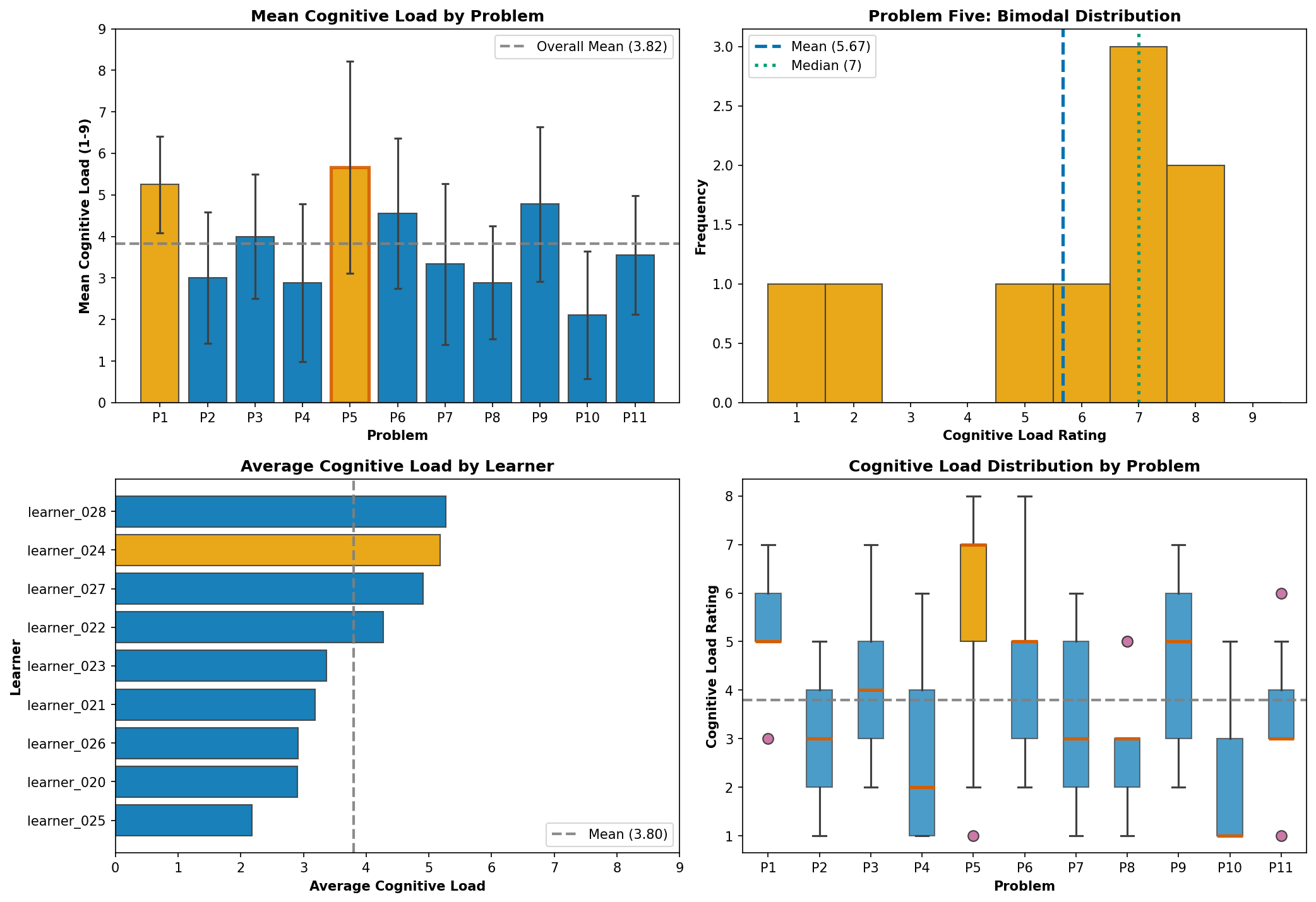}
  \caption{Cognitive load analysis.}
  \Description{}
   \label{fig:f6}
\end{figure*}

Figure \ref{fig:f6} reports the analysis of the Paas scale. This question uses a nine-point Likert scale and asks participants to rate how much mental effort they invested in solving a particular problem from \textit{``very, very low mental effort''} to \textit{``very, very high mental effort.''} Across the eleven problems, the average \textit{within-person} rating (mean of each participant's eleven ratings) was $M=3.80$ ($SD=1.14$, $N=9$), indicating \textit{``rather low mental effort.''} Learner\_028 ($M=5.27$, $SD=0.45$) and Learner\_024 ($M=5.18$, $SD=0.83$) showed uniformly elevated but stable cognitive load across all of the problems. Learner\_026 ($SD=2.35$) and Learner\_021 ($SD=2.12$) showed dramatic swings across the problem set.

Problem five elicited the highest mean cognitive load rating ($M=5.67$, $SD=2.55$, $Mdn=7.0$), followed by problems one ($M=5.25$, $SD=1.16$) and nine ($M=4.78$, $SD=1.86$). The lowest means were for problems ten ($M=2.11$, $SD=1.54$), eight ($M=2.89$, $SD=1.36$), and four ($M=2.89$, $SD=1.90$). Variability was largest on problem five ($SD=2.55$) and smallest on problem one ($SD=1.16$). Although, all problems had $N=9$ ratings except problem one ($n=8$).

Problem five exhibited a bimodal response pattern. Twenty-two percent of learners reported investing very low effort in solving the problems (ratings 1--2) and 56\% of learners reported investing high effort (ratings 7--8). It accounted for 50\% of all high cognitive load ratings ($\ge$7) across the entire dataset. Learner\_021, Learner\_023, and Learner\_026 rated problem five as 7-8 (high difficulty) but their overall average ratings were low (2.91-3.36), suggesting this problem was anomalously difficult for learners who typically found the problems easy. Learner\_025 and learner\_020 found it very easy (1-2), consistent with their overall low-difficulty ratings.

\subsection{Scaffolding Learning Questionnaire Responses}
\begin{table*}
    \caption{Scaffolding Learning Questions}
    \label{tab:scaffolding}
    \centering
    \begin{tabular}{p{0.8\textwidth} l l}
        \toprule
        \textbf{Codespec’s problem solving area:} &  \textit{M}  & \textit{SD} \\
        \midrule
        \rowcolor[HTML]{EFEFEF}
        1. Simplified elements of the computer programming tasks so that they were within my reach. &3.67   &1.32\\
        2. Supported management of the computer programming practice process so that I could engage in elements of programming in realistic contexts (i.e., within an integrated development environment).    &3.89   &0.93\\
        \rowcolor[HTML]{EFEFEF}
        3. Offset my frustration with learning how to program.  &4.33   &0.50\\
        4. Maintained my interest in computer programming. Please Explain. &4.22    &0.67\\
        \rowcolor[HTML]{EFEFEF}
        5. Focused my attention on aspects of the problems that I took for granted. &3.44   &1.01\\
        6. Prompted me to explain why my solutions were successful and to identify the programming concepts I was learning. &3.22   &1.09\\
        \rowcolor[HTML]{EFEFEF}
        7. Prompted me to reflect on my problem-solving process. Please Explain.    &3.78   &0.83\\
        8. Enabled learning by doing in context.    &3.89   &0.93\\
        \bottomrule
    \end{tabular}
    
    \begin{minipage}{0.9\linewidth}
        \footnotesize
        \textit{Notes:} 5-point Likert scale:
        1 = Strongly disagree, 
        2 = Somewhat disagree, 
        3 = Neither agree nor disagree, 
        4 = Somewhat agree, 
        5 = Strongly agree.
    \end{minipage}
\end{table*}
Results showed that, on average, participants \textit{``somewhat agreed''} that Codespec's problem solving area scaffolded their learning.

Participants \textit{``neither agreed nor disagreed''} that Codespec's problem solving area (1) focused their attention on aspects of the problems that they overlooked (item 5) and (2) prompted them to explain why their solution was successful and to identify the programming concepts they were learning (item 6) as shown in Table~\ref{tab:scaffolding}.

The average ratings for two items intended to capture affective, item 3 ($M=4.33$, $SD=0.50$) and item 4 ($M=4.22$, $SD=0.67$) were the highest, and their standard deviation (or $\sigma$) was the lowest.

The one item with particularly high variance in responses gauged how well Codespec's problem solving area simplified elements of the computer programming tasks so that they were within the learners' reach (item 1). Participants also \textit{``somewhat agreed''} that the problem solving area supported management of the computer programming practice process so that they could engage in elements of
programming in realistic contexts (i.e., within an integrated development environment).

Items 4 (about interest) and 7 (about reflection) prompted participants to explain their rating; only six participants responded to the prompt. When explaining why they rated item 4, about the extent to which their interest was maintained by Codespec's problem space area, \textit{``somewhat agree,''} one participant wrote:

\begin{displayquote}
\textit{``I think, in general, interest in programming is correlated with how frustrated one gets in finding a solution to a problem. With Codespec, having those building blocks was a way to offset the frustrating part about programming and allow me to focus on figuring out how to piece those blocks together to achieve the right output. Also, having those blocks acted as a confirmation/checkbox to validate if my original approach was right (which acted as a confidence booster).''} \textemdash Learner\_025.
\end{displayquote}

Other participants who rated item 4 (about interest) \textit{``somewhat agree,''} wrote:

\begin{displayquote}
\textit{``Whenever I ran a program, I got feedback when I was wrong. Based on the test cases, I could focus on where I was wrong, and run the code again and again.''} \textemdash Learner\_023.
\end{displayquote}

\vspace{0.5ex}

\begin{displayquote}
\textit{``Lots of support and resources for programmers on Codespec, such as the Faded Parsons and Pseudocode as well as the test area. Would've liked to see more visual feedback. Links to documentation might've helped me get unstuck and be less frustrated when that happens.''} \textemdash Learner\_026.
\end{displayquote}

\vspace{0.5ex}

\begin{displayquote}
\textit{``It sort of gamifies the coding questions and provide abundant supports/explanations when I made a mistake.''} \textemdash Learner\_027.
\end{displayquote}

\vspace{0.5ex}

\begin{displayquote}
\textit{``The unit tests and pseudocode were especially motivating when solving a problem. They gave me confidence that I could solve the problem if I got stuck.''} \textemdash Learner\_028.
\end{displayquote}



Item 7 (about reflection) prompted participants to explain their rating about the extent to which Codespec's problem solving area prompted them to reflect on their problem-solving process. Most rated item 7 (about reflection) \textit{``somewhat agree.''} One participant wrote:

\begin{displayquote}
\textit{``The two options besides write-code formed a sort of hierarchy in my mind from the high-level overview of the program’s logical structure to the syntax (which could then be converted into code). For example, for the emphasize() program, I wasn’t sure what approach to take to output a string with each word's first letter capitalized. I was considering multiple approaches and possibly overthinking it, so the pseudocode and syntax helped me eliminate some of those more complicated choices and think about the problem at the simplest level.''} \textemdash Learner\_025.
\end{displayquote}

Other participants who rated item 7 (about reflection) \textit{``somewhat agree,''} wrote:

\begin{displayquote}
\textit{``Sometimes the Faded Parsons gave me hints about solving the problems.''} \textemdash Learner\_024.
\end{displayquote}

\vspace{0.5ex}

\begin{displayquote}
\textit{``The output of testcases was clearly presented, guiding me to debug my code.''} \textemdash Learner\_027.
\end{displayquote}

\vspace{0.5ex}

\begin{displayquote}
\textit{``The questions were appropriate for my level, and so it was a good test of the coding knowledge I have acquired so far.''} \textemdash Learner\_028.
\end{displayquote}

One participant who rated item 7 (about reflection) \textit{``strongly agree,''} wrote:

\begin{displayquote}
\textit{``I had a little problem...sometimes I lost confidence, like when solving problem one. But problems two and three were a little bit easy, I believe. When solving those problems, I got my confidence back.''} \textemdash Learner\_023.
\end{displayquote}

One participant who rated item 7 (about reflection) \textit{``neither agreed nor disagreed''} wrote:

\begin{displayquote}
\textit{``I think the availability of tests actually hindered me to some extent because I found myself trying to reverse-engineer the solution from the tests sometimes. At the same time, tests clarified confusing problem descriptions for me. Would've liked to see other solutions or editorials so I could read them after I solve the problems. The pseudocode and faded parsons areas allowed me to piece together solutions sometimes and most importantly get unstuck when I needed.''} \textemdash Learner\_026.
\end{displayquote}
\subsection{Case Study: Problem Five - Whose Left the List?}

\subsubsection{Rationale for Case Study Selection}

Problem five (``Whose Left the List?'') was selected for an in-depth case study analysis because it elicited the highest mean cognitive load rating among the eleven problems ($M=5.67$, $SD=2.55$, $Mdn=7.0$), which varied significantly in perceived difficulty (Friedman $\chi^2(10) = 35.48$, $p < .001$). Problem five accounted for half of all high cognitive load ratings ($\geq7$) in the dataset. Moreover, the standard deviation of 2.55 was nearly double that of the next highest problem, indicating substantial variability in learner experiences.

The distribution of cognitive load ratings for problem five revealed a striking bimodal pattern. While 22.2\% of learners (n=2) reported very low cognitive load (ratings 1-2), and another 22.2\% (n=2) reported moderate load (ratings 5-6), the majority---55.6\% of learners (n=5)---reported high cognitive load (ratings 7-8). This bimodal distribution suggested fundamentally different approaches or understandings among learners, warranting deeper investigation. Of the ten ratings of 7, meaning ``high mental effort,'' or higher across all 98 observations, five of these observations came from problem five alone; this represents 50\% of all such ratings despite being only one of eleven problems.

Additionally, the pattern of cognitive load responses revealed an intriguing paradox: several learners who typically reported low cognitive load across other problems---such as Learner\_021 (overall average 3.18), Learner\_023 (overall average 3.36), and Learner\_026 (overall average 2.91)---rated problem five as exceptionally difficult (7-8), suggesting this problem presented unique challenges that transcended general programming proficiency.

\subsubsection{Problem Five: Task Description and Complexities}

Problem five required learners to implement a function \texttt{locate(lst1,\sloppy lst2)} that would identified elements present in \texttt{lst1} but missing from \texttt{lst2}. The problem statement read:
\vspace{2ex}
\begin{mdframed}
\small
\vspace{2ex}
\textbf{Problem 5 -- Whose Left the List?}

\noindent You are given two lists, \texttt{lst1} and \texttt{lst2}. The elements in \texttt{lst1} have been rearranged, but some of the elements got lost in the process, resulting in \texttt{lst2}. Your task is to write a function called \texttt{locate} that returns a list of the elements that were lost.
\vspace{2ex}

\noindent \textbf{Examples:}
\begin{enumerate}
    \item \texttt{locate([1, 2, 3, 4, 5, 6, 7, 8], [1, 3, 4, 5, 6, 7, 8])} should return \texttt{[2]}
    \begin{itemize}
        \item Explanation: The element 2 is present in \texttt{lst1} but missing in \texttt{lst2}.
    \end{itemize}
    \item \texttt{locate([True, True, False, False, True], [False, True, False, True])} should return \texttt{[True]}
    \begin{itemize}
        \item Explanation: One occurrence of \texttt{True} is missing in \texttt{lst2}.
    \end{itemize}
    \item \texttt{locate([``Jane'', ``is'', ``pretty'', ``ugly''],\allowbreak{} [``Jane'', ``is'', ``pretty''])} should return \texttt{[``ugly'']}
    \begin{itemize}
        \item Explanation: The element \texttt{``ugly''} is present in \texttt{lst1} but missing in \texttt{lst2}.
    \end{itemize}
    \item \texttt{locate([``different'', ``types'', ``5'', 5, [5], [5]],\allowbreak{} [``5'', ``different'', [5], ``types'', 5])} should return \texttt{[[5]]}
    \begin{itemize}
        \item Explanation: One instance of \texttt{[5]} is missing in \texttt{lst2}.
    \end{itemize}
    \item \texttt{locate(list(``rendezvous''), list(``rndvous''))} should return \texttt{[`e', `e', `z']}
    \begin{itemize}
        \item Explanation: The elements \texttt{`e'}, \texttt{`e'}, and \texttt{`z'} are missing in \texttt{lst2}.
    \end{itemize}
    \item \texttt{locate([6, 6, 6, 6, 6, 6, 6, 6, 6], [6, 6, 6, 6, 6, 6])} should return \texttt{[6, 6, 6]}
    \begin{itemize}
        \item Explanation: Three occurrences of 6 are missing in \texttt{lst2}.
    \end{itemize}
\end{enumerate}

\noindent \textbf{Notes:}
\begin{itemize}
    \item Assume that \texttt{lst1} always contains one or more elements.
    \item There will always be elements missing from \texttt{lst1} to result in \texttt{lst2}.
    \item Elements can have duplicates, and some of these duplicates may be lost.
    \item The order of the lost elements in the result list should be the same as their order in \texttt{lst1}.
\end{itemize}
\vspace{2ex}
\end{mdframed}
\vspace{2ex}
\par
The problem incorporated several layers of algorithmic complexity that distinguished it from simpler list manipulation tasks. First, the function needed to handle duplicate elements correctly. If an element appeared three times in \texttt{lst1} but only once in \texttt{lst2}, the function should return that element twice, not three times. Second, the elements could be of any data type---including integers, strings, booleans, lists, dictionaries, floats, and \texttt{None}---requiring learners to implement type-agnostic comparison logic. Third, the order of missing elements in the returned list had to match their original order in \texttt{lst1}, necessitating sequential processing while maintaining awareness of frequency counts.

Test case two exemplifies these complexities. It required learners to recognize that although \texttt{True} appears three times in \texttt{lst1} and twice in \texttt{lst2}, only one occurrence of \texttt{True} is ``missing.'' Furthermore, the elements in \texttt{lst2} are not in the same order as \texttt{lst1}, precluding simple position-based comparison approaches.

The Faded Parsons problem provided as optional scaffolding offered a membership-and-counting solution approach. The code blocks (presented in random order with blanks to fill in) implemented the following logic:

\begin{lstlisting}[language=Python]
def locate(___, lst2):
    missing = []
    for ___ in lst1:
        if i ___ in lst2:
            missing.append(i)
        elif lst1.count(i) ___ lst2.count(i):
            ___.remove(i)
            missing.append(i)
    return ___
\end{lstlisting}

Critically, the Faded Parsons problem included a fix-code block containing a syntax error: \texttt{missing\_append(i)} instead of \texttt{missing. append(i)} (see Figure \ref{fig:f1}). This deliberate bug was designed as a desirable difficulty \cite{Chen2018-xp} to encourage attention to syntactic details. However, as will be discussed, this design choice may have added cognitive load at a point when learners were already struggling with algorithmic understanding.

Additionally, the Faded Parsons problem included two distractor blocks---\texttt{missing = ()} and \texttt{missing.insert(0, i)} (see Figure \ref{fig:f1})---that represented common misconceptions. The first distractor initialized the result as a tuple rather than a list, while the second used an insertion method that would reverse the order of elements.

The Pseudocode Parsons problem provided an alternative scaffolding option, presenting the algorithm in plain English as shown in (see Figure \ref{fig:f2}). This scaffolding made the membership-testing and counting approach explicit through natural language, potentially supporting learners in understanding the algorithmic strategy before confronting syntactic implementation details.


\subsubsection{Learner\_024: Profile and Scaffolding Usage Patterns}

Learner\_024 was selected for a detailed case study analysis because their interaction with problem five (``Whose Left the List?'') exemplified the challenges that emerged when a learner's mental model diverged from the scaffolding's suggested approach. At 37 years old, Learner\_024 was the oldest and most experienced participant in the study, with 13 years of programming experience across JavaScript, HTML/CSS, SQL, Python, Java, C\#, and R. They were pursuing a master's degree in Applied Data Science. Their overall average cognitive load rating across the eleven problems was 5.18 ($SD=1.47$), placing them among the higher cognitive load reporters in the study, though this may reflect the challenge of adapting existing mental models to unfamiliar problem-solving environments rather than fundamental programming difficulties.

Learner\_024's scaffolding usage patterns across the eleven problems revealed selective engagement with the optional scaffolding. They did not view either the Faded Parsons or Pseudocode Parsons problems for problems one, two, three, and seven, suggesting confidence in solving these problems independently. For problem four (``Count in Range''), they briefly viewed both the Pseudocode Parsons and Faded Parsons problems, but did not attempt to solve them, possibly using them for quick reference or confirmation rather than as essential scaffolding. After this first encounter with the Pseudocode Parsons problem during problem four, they only viewed this problem type once more for problem ten, suggesting limited utility for their problem-solving approach. Their engagement with the Faded Parsons problem type increased for later problems; they viewed it for problems six and nine without solving them, attempted to solve the Faded Parsons problem for problem five after unsuccessful write-code attempts, and for problem eleven they both viewed and attempted to solve the Faded Parsons problem (dragging-and-dropping some blocks but not running their solution). This pattern suggests that Learner\_024 sought scaffolding support primarily when encountering significant difficulty, a help-seeking behavior consistent with experienced programmers who prefer independent problem-solving until obstacles necessitate external resources.

Learner\_024's preference for Faded Parsons over Pseudocode Parsons problems was explicitly articulated in their retrospective interview. When asked about their experiences with the scaffolding options, they explained:

\begin{displayquote}
\textit{``Pseudocode, maybe helpful, but not for me because I'd have to understand the pseudocode text and switch to my code and that looked like a heavy or difficult task. I had limited time to understand the pseudocode text. I used the Faded Parsons problems because they gave me hints about coding because it described things in Python code, so I was able to directly understand the code I wrote and the Faded Parsons problem's code....I didn't consider all of the Python sentences. I just had to fill-in-the blanks when I use the Faded Parsons, so it was really helpful.''} \textemdash Learner\_024.
\end{displayquote}

This statement reveals that Learner\_024 perceived the natural language descriptions in Pseudocode Parsons problems as an additional translation burden---requiring cognitive effort to convert English descriptions into Python syntax---while the partially completed Python code in Faded Parsons problems could be directly compared to their own code. The preference for syntax-level hints over algorithmic descriptions suggests that Learner\_024's struggle with problem five was primarily conceptual (understanding which algorithmic approach to use) rather than syntactic (knowing how to express an algorithm in Python). Notably, Learner\_024's first language appeared to be Japanese based on interface language settings captured during the session, which may have contributed to the additional cognitive load of processing English-language algorithmic descriptions under time pressure, though this factor was not explicitly mentioned in their explanation.

\subsubsection{Five Attempts: Mental Model Mismatch and Scaffolding Divergence}

Learner\_024 made five total attempts at problem five: three write-code attempts, two subsequent edits to the write-code solution after viewing scaffolding, and two attempts at the Faded Parsons problem. Each phase of attempts revealed distinct characteristics of their problem-solving approach and the challenges they encountered.

\paragraph{Write-Code Attempts 1-3: The Sequential Matching Approach}

Learner\_024's initial approach to problem five employed what can be characterized as a sequential matching strategy. Their first write-code attempt, submitted approximately 2 minutes into the problem-solving period, implemented the following logic:

\begin{lstlisting}[language=Python]
def locate(lst1, lst2):
    j = 0
    l = []
    for i in range(len(lst1)):
        if lst1[i] == lst2[j]:
            j += 1
            continue
        else:
            l.append(lst1[i])
    return l
\end{lstlisting}

This solution reveals Learner\_024's mental model for the problem: they conceptualized finding missing elements as a process of stepping through \texttt{lst1} position by position, checking whether each element matched the current position in \texttt{lst2}. When elements matched, they advanced the \texttt{lst2} pointer (\texttt{j}); when there was no match, Learner\_024 concluded the element was missing. This approach assumes that \texttt{lst2} maintains some sequential relationship to \texttt{lst1}, with elements appearing in the same relative order even if some are missing.

The solution failed immediately due to an \texttt{IndexError: list index out of range}. The code did not verify whether \texttt{j} remained within the bounds of \texttt{lst2} before accessing \texttt{lst2[j]}, causing the program to crash when \texttt{j} exceeded the length of \texttt{lst2}.

Learner\_024 began their second attempt 10 seconds after submitting the first and worked on it for 29 seconds before submitting. This attempt attempted to address the issue by adding a boundary check:

\begin{lstlisting}[language=Python]
def locate(lst1, lst2):
    j = 0
    l = []
    for i in range(len(lst1)):
        if len(lst2) < j and lst1[i] == lst2[j]:
            j += 1
            continue
        else:
            l.append(lst1[i])
    return l
\end{lstlisting}

However, the boundary condition was incorrect: \texttt{len(lst2) < j} would only be \texttt{True} when \texttt{j} exceeded the list length, which would already cause an index error. The condition should have been \texttt{len(lst2) > j} to verify that accessing \texttt{lst2[j]} was safe. This attempt failed with the same error.

Learner\_024 began their third attempt 33 seconds after submitting their solution for the second attempt and worked on it for only 5 seconds before submitting. This attempt corrected the boundary condition:

\begin{lstlisting}[language=Python]
def locate(lst1, lst2):
    j = 0
    l = []
    for i in range(len(lst1)):
        if len(lst2) > j and lst1[i] == lst2[j]:
            j += 1
            continue
        else:
            l.append(lst1[i])
    return l
\end{lstlisting}

This version executed without crashing and passed eight of 10 test cases. However, it failed two critical tests\textemdash three and four (see the problem statement). These failures reveal the fundamental limitation of the sequential matching approach: it cannot handle cases where \texttt{lst2} contains the same elements as \texttt{lst1} but in a different order. The algorithm assumes that if an element in \texttt{lst1} does not match the current position in \texttt{lst2}, that element must be missing---but in reality, the element might simply appear later in \texttt{lst2}. For test three, the algorithm incorrectly identified multiple elements as missing because their positions in \texttt{lst2} differed from their positions in \texttt{lst1}.



\paragraph{Transition to Scaffolding: Attempted Integration}

After the third write-code attempt failed tests three and four, Learner\_024 viewed the Pseudocode Parsons problem tab briefly (less than 2 seconds) before switching to the Faded Parsons problem. Despite their stated preference for Faded Parsons problems due to this problem's direct presentation of Python syntax, Learner\_024 did not immediately attempt the Faded Parsons problem. Instead, they returned to their write-code solution and made two additional edits over the next approximately 55 seconds, apparently attempting to incorporate elements of the scaffolding approach into their existing sequential matching framework. This behavior suggests a reluctance to abandon their original approach entirely, consistent with their perception that the Faded Parsons problems provided ``hints'' rather than alternative algorithmic strategies.

The first edit added two lines, but did not fundamentally alter the logic:

\begin{lstlisting}[language=Python]
def locate(lst1, lst2):
    j = 0
    l = []
    d = len  # New line - Incomplete thought
    for i in range(len(lst1)):
        if len(lst2) > j and lst1[i] == lst2[j]:
            j += 1
            continue
        else:
            l.append(lst1[i])
    return l
\end{lstlisting}

The line \texttt{d = len} appears to be an incomplete thought, possibly an attempt to create a helper variable for length checking. It was not used elsewhere in the code and did not affect execution. This edit suggests Learner\_024 recognized a need to modify their approach, but struggled to identify what specifically needed to change.

The second edit reveals a more substantive attempt to incorporate the \texttt{count()} method, which was central to the Faded Parsons solution:

\begin{lstlisting}[language=Python]
def locate(lst1, lst2):
    j = 0
    l = []
    for i in range(len(lst1)):
        e = len(lst1[i])  # New line
        lst2.count(e)      # New line
        if len(lst2) > j and lst1[i] == lst2[j]:
            j += 1
            continue
        else:
            l.append(lst1[i])
    return l
\end{lstlisting}

Here, Learner\_024 attempted to use \texttt{lst2.count()}, but the implementation reveals a misunderstanding of how the counting approach should work. First, \texttt{e = len(lst1[i])} would only work if \texttt{lst1[i]} were itself a sequence (list, string, etc.), and even then, it would count the length of the element rather than the element itself. Second, the result of \texttt{lst2.count(e)} was not stored or used in any conditional logic---the function was called, but its return value was discarded. This suggests Learner\_024 recognized that \texttt{count()} was relevant to solving the problem but could not integrate it meaningfully into their sequential matching framework.

These two edits consumed approximately 55 seconds. Given that the total problem-solving time was 10 minutes and Learner\_024 had already spent approximately 2 minutes on their first write-code attempt, 29 seconds on their second, and 5 seconds on their third, they had used roughly 8 minutes and 4 seconds before beginning the Faded Parsons attempts, leaving approximately 1 minute and 56 seconds for the two Faded Parsons attempts.

\paragraph{Faded Parsons Attempts 1-2: Active Rejection of the Counting Approach}

Learner\_024 made two attempts at the Faded Parsons problem during the final approximately 1 minute and 56 seconds of the problem-solving period. Both attempts revealed the same two errors: a parameter name mismatch and failure to correct the fix-code syntax bug. However, screen recording analysis revealed a critical detail about their engagement with the scaffolding's algorithmic approach that warrants careful interpretation.

The first Faded Parsons attempt, submitted with 25 seconds remaining on the clock was the following:

\begin{lstlisting}[language=Python]
def locate(l, lst2):  # Should be lst1, not l
    missing = []
    for i in lst1:
        if i not in lst2:
            missing_append(i)  # Syntax error: should be missing.append(i)
    return missing
\end{lstlisting}

Learner\_024 correctly filled in several blanks; they identified that \texttt{missing} should be initialized as an empty list \texttt{[]}, that the loop variable should be \texttt{i}, and that the membership test should be \texttt{not in}. However, they made two critical errors. First, they typed \texttt{l} instead of \texttt{lst1} as the first parameter name, creating a mismatch between the parameter (\texttt{l}) and its usage within the function body (\texttt{lst1}). Second, and most significantly, they did not correct the fix-code block's syntax error: \texttt{missing\_append(i)} should be \texttt{missing.append(i)}. The underscore should be a dot operator to call the \texttt{append} method on the \texttt{missing} list object.

Critically, screen recording analysis revealed that Learner\_024 did not simply omit the \texttt{elif} branch---they actively engaged with it and then rejected it. During the construction of their first attempt, they dragged the \texttt{elif lst1.count(i) \_\_\_ lst2.count(i):} block from the available blocks area and positioned it beneath the \texttt{if i not in lst2:} block. However, before submitting, they moved the block back out of their solution. This behavior suggests that Learner\_024 recognized the counting logic as potentially relevant but made a deliberate choice to exclude it. Several interpretations are possible. Learner\_024 may not have known how to fill in the comparison operator blank (the \texttt{\_\_\_} requiring a \texttt{>} symbol), they may have wanted to test the simpler membership-only approach first, or the added complexity may have felt overwhelming given the severe time pressure. Regardless of the specific reason, this represents active rejection after consideration rather than simple oversight or omission due to time constraints.

The second Faded Parsons attempt, submitted with only 10 seconds remaining, corrected the parameter name but retained the syntax error:

\begin{lstlisting}[language=Python]
def locate(lst1, lst2):  # Corrected
    missing = []
    for i in lst1:
        if i not in lst2:
            missing_append(i)  # Syntax error remains
    return missing
\end{lstlisting}

This suggests Learner\_024 noticed and corrected the parameter mismatch, but they failed to recognize the incorrect method call syntax. The \texttt{elif} branch remained excluded. With only 10 seconds between their first and second submissions, Learner\_024 had no realistic opportunity to reconsider the counting approach they had previously rejected.

\subsubsection{Mental Model Analysis: Sequential Matching vs. Membership Testing}

The divergence between Learner\_024's approach and the Faded Parsons solution represents a fundamental difference in mental models for the problem. Learner\_024's sequential matching model conceptualized the task as: ``Step through lst1 position by position, checking if each element matches the current position in lst2. If not, the element is missing.'' This model assumes that lst2 preserves the relative ordering of elements from lst1, even if some elements are removed.

In contrast, the Faded Parsons membership-and-counting model conceptualized the task as: ``For each element in lst1, check if it exists anywhere in lst2. If not, it's missing. If it exists but appears fewer times in lst2 than in lst1, then some occurrences are missing.'' This model makes no assumptions about element order and instead relies on set membership and frequency counting.

The two mental models are not merely different implementation strategies for the same high-level approach---they represent fundamentally different understandings of what the problem is asking. Sequential matching treats the problem as a synchronization task that aligns two sequences and identifies where they diverge. Membership testing reframes the problem; rather than tracking what moved, it counts how many of each element appear in both lists and identifies the shortfall.

Critically, Learner\_024's attempts to integrate the \texttt{count()} method into their sequential matching framework suggest they recognized that counting was relevant but could not reconcile it with their positional iteration approach. The \texttt{count()} method is most naturally used when iterating over elements (\texttt{for i in lst1}) rather than over indices (\texttt{for i in range(len(lst1))}), because it asks ``how many times does this element appear?'' rather than ``what element is at this position?'' Learner\_024's hybrid approach---iterating over indices while trying to use \texttt{count()}---created cognitive dissonance that manifested in nonsensical code like \texttt{e = len(lst1[i]); lst2.count(e)}.

The screen recording, which provides evidence that Learner\_024 physically positioned and then removed the \texttt{elif} counting block, adds important nuance to this analysis. Their rejection was not due to failure to notice the counting approach---they directly engaged with it. Rather, the rejection suggests they could not reconcile this approach with their understanding of the problem within the available time, or could not determine how to complete the block (specifically, what comparison operator to use). This active engagement followed by rejection illustrates the cognitive challenge of abandoning a familiar mental model. Even when presented with an alternative approach in concrete syntactic form, learners may recognize its relevance without being able to integrate it into a coherent solution.

This mental model mismatch created a situation where the Faded Parsons problem's scaffolding, rather than reducing cognitive load, may have increased it. Instead of providing a ladder to reach a solution along the path Learner\_024 was already following, the scaffolding presented an entirely different path. Learner\_024 was left in the cognitively demanding position of either abandoning their approach entirely to adopt the scaffolding's approach (which required understanding why the membership approach was correct and their approach was flawed), or attempting to reconcile the two approaches (which led to the unsuccessful hybrid attempts). With only approximately 1 minute and 56 seconds remaining when they began the Faded Parsons problem attempts, neither option was realistic.

\subsubsection{Questionnaire Responses: Surface-Level Engagement}

Learner\_024's responses to the Scaffolding Learning Questionnaire provide additional insight into their experience with the optional scaffolding. Their average rating across the eight items was 3.75 ($SD=0.46$), indicating that they ``somewhat agreed'' that Codespec's problem space area effectively scaffolded their learning. A closer examination of how the behavioral data from Learner\_024's problem-solving attempts aligned with their questionnaire responses shows the scaffolding succeeded in maintaining their engagement---Learner\_024 persisted through five attempts and used the available resources rather than giving up---but it did not help them understand the differences between their approach and the Faded Parsons and or Pseudocode Parsons problem's solution. Programmers may also be less likely to question their initial approach to a problem under time pressure.

Learner\_024 rated the following items ``somewhat agree'': item 1 (``simplified elements of computer programming tasks so that they were within my reach''), item 2 (``supported management of the computer programming practice process''), item 3 (``offset my frustration with learning how to program''), and item 8 (``enabled learning by doing in context''). These ratings suggest that the availability of scaffolding options helped Learner\_024 feel supported and maintained their engagement with challenging problems.

However, their rating for item 4 (``maintained my interest in computer programming'') was only 3 (``neither agree nor disagree''), and their open-ended explanation revealed a more complex perspective:

\begin{displayquote}
\textit{``It looks scratch and it is interesting, but I've created many programs and my interest was not changed.''} \textemdash Learner\_024.
\end{displayquote}

This response suggests that as an experienced programmer with 13 years of practice, Learner\_024's interest in programming was already stable and not significantly influenced by the scaffolding environment. The visual, block-based interface (which they compared to Scratch) was novel, but it did not fundamentally alter their interest with programming as a discipline. Additionally, in the retrospective interview, Learner\_024 said, ``I was not motivated because coding is required for work, so my attitude to the programming was not changed.''

More revealing are Learner\_024's ratings and responses for items intended to capture metacognitive processes. They rated item 5 (``focused my attention on aspects of problems that I took for granted'') as only 3 (``neither agree nor disagree''), and item 6 (``prompted me to explain why my solution was successful and to identify the programming concepts I was learning'') as 4 (``somewhat agree''). For item 7 (``prompted me to reflect on my problem-solving process''), Learner\_024 gave a rating of 4 and provided this explanation:

\begin{displayquote}
\textit{``Sometimes the Faded Parsons gave me hints about solving the problems.''} \textemdash Learner\_024.
\end{displayquote}

This explanation characterizes the Faded Parsons problems as providing ``hints'' rather than prompting reflection on problem-solving strategies. The phrasing suggests a surface-level engagement with the scaffolding---viewing it as a source of implementation details rather than as a prompt to reconsider their algorithmic approach. Notably, Learner\_024 did not mention the mental model mismatch or recognize that the Faded Parsons problem solution employed a fundamentally different strategy than their sequential matching approach. This lack of metacognitive awareness may explain why their attempts to integrate \texttt{count()} into their existing framework were unsuccessful; they treated it as a code snippet to incorporate (a ``hint'') rather than as an indicator that a different conceptual approach was needed.

\subsubsection{Connections to Research Questions}

The case study of Learner\_024's engagement with problem five provides concrete evidence addressing both research questions. The first research question (\hyperref[rq:rq1]{RQ1}) asked, ``What do learners perceive are the advantages and challenges of using Faded Parsons and Pseudocode Parsons problems as optional scaffolding for write-code problems?'' The case study analysis revealed a critical challenge. When the scaffolding's solution diverges from the learner's mental model, the scaffolding may increase rather than decrease cognitive load. Learner\_024's attempts to reconcile their sequential matching approach with the membership-and-counting approach presented in the Faded Parsons problem consumed valuable time and resulted in unsuccessful hybrid solutions. The screen recording evidence is particularly illuminating here. Learner\_024 physically dragged the \texttt{elif} counting block into position, demonstrating direct engagement with the scaffolding's algorithmic approach, but then deliberately removed it before running their code. This active rejection---rather than passive omission---suggests that scaffolding can successfully draw attention to alternative approaches without providing sufficient support for learners to understand why that approach is necessary or how to implement it correctly. The challenge was compounded by the inclusion of a fix-code syntax bug (see Figure \ref{fig:f1}), which added an additional debugging task at precisely the moment when Learner\_024 was already struggling to evaluate competing algorithmic strategies.

However, the case study analysis also revealed advantages. The availability of scaffolding maintained Learner\_024's engagement despite repeated failures---they continued attempting the problem rather than moving on or giving up. Their questionnaire responses indicate that the scaffolding helped offset frustration, even if it did not resolve the underlying algorithmic confusion. Their explicit preference for Faded Parsons over Pseudocode Parsons problems---valuing direct Python code over natural language descriptions that required translation---reveals that learners make strategic choices about which scaffolding types align with their problem-solving approach and cognitive needs. This suggests that scaffolding can serve important affective functions even when it falls short of its cognitive objectives, and that providing multiple scaffolding options allows learners to select formats that minimize extraneous cognitive load.

The second research question (\hyperref[rq:rq2]{RQ2}) asked, ``What preliminary insights emerge from piloting the Scaffolding Learning Questionnaire with novice programmers using Codespec?'' The case study analysis illustrates both the questionnaire's strengths and its limitations. The questionnaire successfully captured Learner\_024's sense of being supported\textemdash their ratings of 4 (``somewhat agree'') for items addressing task simplification, frustration offset, and realistic engagement aligned with behavioral evidence of persistence through five attempts. However, the items intended to capture metacognitive processes revealed a more complex picture. While Learner\_024 rated items 6 and 7 (explaining solutions, reflecting on problem-solving) as 4, their qualitative explanation\textemdash characterizing the Faded Parsons problems as providing hints\textemdash suggests this reflection remained at a surface level. This aligns with their observed behavior of engaging with individual code blocks (like the \texttt{elif} counting block) without stepping back to consider why the scaffolding's overall approach differed from their own. Notably, item 5 (``focused my attention on aspects of problems that I took for granted'') received a rating of only 3, which also aligns with the behavioral evidence from screen recordings. Learner\_024 did not recognize the fundamental difference between their sequential matching approach and the scaffolding's membership-and-counting strategy. This suggests the questionnaire's open-ended prompts were valuable for revealing the quality of metacognitive engagement that numerical ratings alone would obscure. Future iterations might probe more deeply into learners' awareness of mental model mismatches by explicitly asking whether the scaffolding suggested a different approach than they had initially considered, and if so, how they responded to that difference.

Additionally, the case study highlights the role of individual differences in scaffolding effectiveness. Learner\_024's 13 years of programming experience, while making them proficient at implementing algorithms once conceptualized, may have also made them less flexible in reconsidering their initial problem formulation. Learner\_024's active rejection of the counting approach---despite physically engaging with the relevant code block---suggests that experienced programmers may evaluate scaffolding against their existing mental models and dismiss approaches that do not immediately align, rather than treating the scaffolding as a signal to reconsider their understanding of the problem. Scaffolding design that assumes learners will readily adopt a presented approach may be less effective for experienced programmers with established problem-solving habits. This finding has implications for adaptive scaffolding systems. Rather than assuming that more detailed scaffolding is always better, systems might need to detect when a learner's approach diverges from the scaffolding and provide explicit metacognitive prompts like ``The solution shown uses a different approach than the one you tried. Can you identify how they differ?'' or ``This block handles cases where an element appears in both lists but with different frequencies. When might that matter?''

Finally, the case study underscores the importance of time-\sloppy management in scaffolded problem-solving and the fact that time-pressured problem-solving environments may systematically disadvantage reflective learners \cite{Phillips2016-ka}. Learner\_024 spent approximately 8 minutes and 4 seconds on their write-code attempts and subsequent edits before turning to the Faded Parsons problem scaffolding, leaving only approximately 1 minute and 56 seconds to understand and implement a different approach. Within that constrained window, they engaged with the counting block, rejected it, and made two attempts---the second submitted with only 10 seconds remaining. An adaptive system might intervene earlier, detecting when a learner is making multiple unsuccessful attempts with the same flawed approach and prompting them to consider alternative strategies before time pressure becomes overwhelming. The system might also detect when a learner engages with a scaffolding element (such as dragging a block into position) and then removes it, using this as a signal to provide targeted support explaining that element's purpose.
\section{Discussion}
This study explored the perceived advantages and challenges of interacting with a new computer programming practice environment, called Codespec, and its ability to scaffold learning when learners are given optional scaffolding\textemdash the choice of viewing and or solving Faded and Pseudocode Parsons problems to help them complete write-code problems. Programmers were asked to solve eleven computer programming problems, rate how much mental effort they invested in doing so, complete a retrospective think-aloud interview, and respond to a questionnaire about the ability of Codespec's problem solving area to scaffolding learning.

\subsection{Time Pressure and Reflective Processing}
\label{sec:time}
According to the results, an advantage of offering both Faded and Pseudocode Parsons problems as optional scaffolds is that they can support learners' progress on write-code tasks by enabling comprehension monitoring, encouraging strategy formation, and helping to refine existing knowledge. Several learners described writing an initial solution and then using either scaffold to verify logic or syntax, indicating that these problem types function as externalized checkpoints that help learners calibrate understanding while reducing uncertainty. This is consistent with prior research that found learners use Parsons problems to debug their solution, and that the benefits of using Parsons problems to scaffold write-code problems includes having the option to (1) scan the blocks without solving it, and (2) solve it and replace or modify their write-code solution \cite{Hou2022-xa}. This post-completion use of scaffolding for verification represents what Loksa et al. \cite{Loksa2020-cy} term ``evaluating''\textemdash assessing whether one's solution produces the expected results. Notably, this was one of the two self-regulation behaviors that all participants in Loksa et al.'s study exhibited, suggesting it may be a foundational metacognitive skill that scaffolding environments can productively support.

However, the results also reveal an important boundary condition. Prior work found that using adaptive Parsons problems to scaffold write-code problems can also reduce difficulty and completion time when participants are allowed to skip problems in a problem set or stop at any time within a set duration \cite{Hou2022-xa}. In contrast, the results of this study show efficiency benefits may not generalize to contexts where each problem is individually timed. Under per-problem time constraints, several learners delayed seeking scaffolding until the final minutes\textemdash Learner\_024, for instance, spent approximately eight minutes on write-code attempts before turning to the Faded Parsons problem, leaving insufficient time to understand and implement an alternative approach. This pattern is consistent with Phillips et al.'s \cite{Phillips2016-ka} meta-analysis finding that time pressure weakens the relationship between reflective thinking and decision performance, suggesting that timed problem-solving environments may systematically limit learners' capacity to benefit from scaffolding that requires reconceptualizing their approach.

This distinction between problem-set timing and per-problem timing has implications for how we design environments that support optional scaffolding in both summative and formative assessment contexts. For summative assessments where per-problem timing may be necessary, systems could provide approach-based feedback indicating how closely a learner's current solution aligns with the scaffolding, enabling faster decisions about whether to change strategies. For formative practice contexts, Codespec supports instructors in allowing learners to pause a problem, resulting in the interface hiding the problem and displaying a button that reads `Resume' alongside the message ``Click 'resume' to continue.'' This pause functionality preserves time for reflection without penalizing learners who need additional processing time\textemdash though future research should examine whether learners actually use this feature and whether it mitigates help-avoidance behaviors.

\subsection{Faded Parsons Problems as a Desirable Difficulty}
According to the results, two participants preferred Faded Parsons problems specifically because the fill-in-the-blanks made them more challenging. Learner\_030 explained: ``I think the Faded Parsons problem is interesting because it doesn't give you all the answers and it helps you see what variables are necessary and how you could structure your code to arrive at a solution.'' This perception of Faded Parsons problems as a desirable difficulty \cite{Chen2018-xp} extends Weinman, Fox, and Hearst's \cite{Weinman2021-in} finding that learners preferred Faded Parsons over code-writing questions. This study's results suggest some learners also prefer Faded Parsons over standard Parsons problems because the partial completion requirement maintains productive challenge.

However, this finding also revealed an important design oversight Fade Parsons problems and Codespec's implementation. Most systems that support Faded Parsons problems do not provide a help feature for filling in the blanks, leaving learners who get stuck on a blank without a way to progress. This may explain why some learners might find the scaffolding less useful than expected.

The implication for design is that scaffolding systems supporting Faded Parsons problems should include progressive hints for blank completion\textemdash for example, revealing the type of expression needed, then providing a partial fill-in, and finally revealing the complete code. This would preserve the desirable difficulty while preventing learners from becoming stuck at an intermediate scaffolding level.

\subsection{Time-on-Task and Mental Model Mismatch}
According to the results, learners reported two primary challenges when using Faded and Pseudocode Parsons problems as scaffolding: the time required for drag-and-drop interaction and confusion when the scaffolding solution diverged from their mental model. Regarding time cost, Learner\_022 explained: ``I'm not familiar with debugging or figuring things out with Faded and Pseudocode Parsons problems. I felt that they were a bit confusing and a bit complex to use because you needed to drag-and-drop the blocks.'' Learner\_021 similarly noted that ``it would take longer for me to drag-and-drop them.'' Furthermore, the behavioral evidence from the case study analysis showed the the fix-code block containing a syntax error may compound these challenges. These concerns echo prior findings that drag-and-drop interaction, in particular, with distractor blocks, adds to time-on-task, though the efficiency cost must be weighed against the potential learning benefits \cite{Smith2023-pv, Smith2024-cz}.

More consequentially, learners experienced confusion when the scaffolding's approach did not align with their existing mental model. Learner\_025 described this challenge saying, ``supposing I had some idea already in my head, but it wasn't super clear, and I wanted to look over the options for the blocks in the Pseudocode problem. If some of the blocks did not correspond to the idea that I had in my head for the program, it would get a bit confusing.'' This finding is consistent with Hou, Ericson, and Wang's \cite{Hou2022-xa} observation that ``some students found [Parsons problems] less useful if the Parsons solution did not match their approach or if they did not understand the solution.''

The case study analysis of Learner\_024's interaction with problem five illuminates the cognitive mechanisms underlying this challenge. Screen recordings revealed that Learner\_024 physically positioned the \texttt{elif} counting block into their solution and then deliberately removed it before submitting\textemdash demonstrating active engagement with the scaffolding's approach followed by rejection. This behavior suggests that the difficulty is not simply a preference for one's own approach, but rather a failure to recognize that the scaffolding represents a fundamentally different algorithmic strategy. Learner\_024's sequential matching approach and the scaffolding's membership-and-counting approach were not variations on a single strategy but distinct mental models of the problem itself.

This pattern connects to Loksa et al.'s \cite{Loksa2020-cy} finding that novice programmers rarely engage in ``interpreting'' (understanding the problem at a conceptual level) and ``adapting'' (modifying prior solutions to fit new contexts). These were the two least-exhibited self-regulation behaviors in their study, with only 32\% of participants demonstrating either behavior. The present study extends this finding by showing that even experienced programmers may fail to interpret when a scaffolding solution requires a different conceptual approach\textemdash particularly under time pressure. Learner\_024's 13 years of programming experience did not prevent them from treating the scaffolding as a source of syntactic ``hints'' rather than as a signal to reconsider their algorithmic approach.

Reiser and Tabak \cite[cf.,][p. 61]{Reiser2014-ql} argue that scaffolding can serve a ``problematizing'' function\textemdash helping learners attend to aspects of problems they might otherwise take for granted. However, this function requires that learners recognize the discrepancy between their approach and the scaffolding's approach. When learners lack this metacognitive awareness, the scaffolding may add cognitive load without providing corresponding benefit. The implication for design is that scaffolding systems should explicitly surface approach-level differences. For example, when a learner views a Faded Parsons problem after attempting a write-code solution, the system could highlight: ``Your solution iterates through indices using \texttt{range(len(lst1))}. The scaffolding iterates through elements directly using \texttt{for i in lst1}. These approaches handle element order differently\textemdash consider when this might matter.''

\subsection{Strategic Scaffolding Selection}
According to the results, learners selectively used Faded Parsons problems for syntax and structure support while using Pseudocode Parsons problems for high-level reasoning. Learner\_026 explained, ``I used to teach high schoolers computer programming and the part they struggled with was not the syntax, but mostly the pseudocode itself\textemdash the problem-solving process itself....I would keep Pseudocode Parsons for beginners instead of Faded Parsons.'' Conversely, Learner\_024 preferred Faded Parsons problems because ``they gave [them]] hints about coding, [and] it described things in Python code, so [they were] able to directly understand the code I wrote and the Faded Parsons problem's code.''

This strategic differentiation extends prior work on Faded Parsons problems \cite{Weinman2021-in} and subgoal labels \cite{Morrison2016-qd} by demonstrating that learners perceive these scaffolding types as serving complementary functions. The finding also aligns with Reiser and Tabak's \cite[cf.][]{Reiser2014-ql} concept of \textit{differentiated scaffolding}, in which ``different means [are used] to support different aspects of performance.'' The results suggest that offering multiple scaffolding types allows learners to select support that matches their current needs---whether syntactic (Faded Parsons) or conceptual (Pseudocode Parsons). Future research should also explore incorporating other research-based problem types such as fix-code problems \cite{Ericson2017-zo} and or the more recent Prompt problems \cite{Denny2024-kx} as well as prompt learners to develop examples/test cases \cite{Pechorina2023-kd, Wrenn2019-hp}.

\subsection{Toward an Instrument for Measuring the Effectiveness of Scaffolds}
This study piloted the Scaffolding Learning Questionnaire, an eight-item instrument derived from Reiser and Tabak's \cite{Reiser2014-ql} six aspects of scaffolding. The questionnaire yielded preliminary insights about how learners perceive scaffolding, but also revealed measurement challenges that future research should address.

The response pattern showed higher ratings and lower variance for affective items (item 3, frustration offset: $M = 4.33$, $SD = 0.50$; item 4, interest maintenance: $M = 4.22$, $SD = 0.67$) compared to metacognitive items (item 5, attention to overlooked aspects: $M = 3.44$, $SD = 1.01$; item 6, explanation and concept identification: $M = 3.22$, $SD = 1.09$). This pattern could reflect genuine differences in scaffolding effectiveness across functions, but it could also reflect differences in item interpretability. Affective states like frustration and interest may be more accessible to self-report than metacognitive processes like attention allocation and self-explanation, which learners may not consciously monitor.

The qualitative responses revealed a related issue: learners interpreted the same scaffolding features through different cognitive frames. Learner\_025 described the scaffolding options as forming ``a sort of hierarchy in my mind,'' while Learner\_024 characterized them as providing ``hints.'' These framings suggest different mental models of what scaffolding is and how it should be used---yet both learners might rate the same questionnaire items similarly despite fundamentally different experiences.

This measurement challenge reflects a broader problem in computing education research. Zavaleta Bernuy and Harrington's systematic review of student surveys across 15 years of SIGCSE, ITiCSE, and ICER publications revealed that only four topics have been covered demographics, abilities and experiences, perceptions of computer science, and motivation \cite{Zavaleta-Bernuy2020-ra}.

The implication for research is that evaluating scaffolding effectiveness may require triangulating self-report measures with behavioral data. In this study, Learner\_024's questionnaire responses indicated that scaffolding ``sometimes gave hints,'' yet screen recordings revealed active engagement with and rejection of the scaffolding's algorithmic approach---a pattern invisible to self-report. Future iterations of the Scaffolding Learning Questionnaire should include items that probe learners' awareness of approach-level differences between their solution and the scaffolding, and should be validated against behavioral measures of scaffolding use.

\subsection{Limitations}

This study has several limitations that should be considered when interpreting the findings and that suggest directions for future research.

First, the sample size ($N = 9$) limits the generalizability of the results. While this sample was appropriate for an exploratory, qualitative focus---particularly for the retrospective think-aloud interviews and case study analysis---it constrains the statistical power of quantitative measures such as the Paas scale ratings and Scaffolding Learning Questionnaire responses. The correlational patterns and between-problem comparisons reported should be interpreted as preliminary observations rather than definitive findings.

Second, the sample composition was homogeneous along several dimensions. Eight of nine participants identified as male and eight identified as Asian; five were Computer Science majors and three were in related fields (Electrical Engineering and Computer Science, Applied Data Science, Educational Technology). Furthermore, participants reported between 1 and 13 years of programming experience ($M = 56.33$ months); they were not truly novice programmers in that they had extensive experience with other programming languages. This experienced sample may have exhibited different scaffolding preferences and usage patterns than would true novices taking their first programming course. The finding that learners selectively used scaffolding and sometimes rejected it entirely may not generalize to learners without established problem-solving habits and mental models.

Third, the single-session design precluded assessment of long-term learning outcomes or transfer effects. While prior research has demonstrated that Parsons problems and Faded Parsons problems facilitate learning transfer~\cite{Weinman2021-in, Smith2024-cz}, this study focused on immediate problem-solving behavior and self-reported perceptions. We cannot determine whether the scaffolding strategies learners employed led to durable improvements in programming skill or conceptual understanding.

Fourth, the per-problem time constraint of ten minutes per problem created conditions that may not reflect typical learning environments. As discussed in Section \ref{sec:time}, time pressure has been shown to weaken the relationship between reflective thinking and decision performance \cite{Phillips2016-ka}. The observation that learners delayed seeking scaffolding until late in the problem-solving period---leaving insufficient time to benefit from alternative approaches---may be an artifact of the timed protocol rather than a general characteristic of scaffolding use. Future studies should compare scaffolding engagement under timed versus untimed conditions.

Fifth, this study lacked a control condition, precluding causal claims about scaffolding effectiveness. All participants had access to both Faded Parsons and Pseudocode Parsons problems as optional scaffolds; we cannot determine whether the availability of these scaffolds improved problem-solving performance relative to a write-code-only condition. The observed benefits (comprehension monitoring, strategy formation, frustration offset) are based on learners' self-reports rather than controlled comparisons.

Sixth, the retrospective think-aloud methodology, while valuable for capturing learners' perceptions and reasoning, is subject to recall limitations and post-hoc rationalization. Participants reflected on their problem-solving experiences after completing all eleven problems, potentially introducing memory distortions or socially desirable responses. Additionally, although the researcher and an undergraduate research assistant achieved 100\% agreement on coding, the deductive coding approach based on prior work~\cite{Hou2022-xa} may have constrained the identification of novel themes specific to Faded and Pseudocode Parsons problems.

Seventh, the Scaffolding Learning Questionnaire has not yet been validated. While the items were derived from Reiser and Tabak's \cite{Reiser2014-ql} theoretical framework for scaffolding, this study represents only a pilot administration. The observed pattern of higher ratings for affective items compared to metacognitive items may reflect genuine differences in scaffolding effectiveness, differences in item interpretability, or some combination thereof. Future work should establish the questionnaire's psychometric properties through factor analysis and convergent validity studies with behavioral measures.

Eighth, although screen recordings captured learner behavior during problem-solving, the case study analysis focused intensively on a single learner (Learner\_024) and a single problem (problem five). While this depth of analysis revealed important insights about mental model mismatch and scaffolding rejection, the extent to which these patterns characterize other learners' experiences remains unclear. A more comprehensive analysis of screen recordings across all participants would strengthen the findings.

Finally, the study examined only two scaffolding variations---Faded Parsons and Pseudocode Parsons problems---while Codespec supports additional problem types including fix-code problems and regular Parsons problems. The relative effectiveness and learner preferences for a broader range of scaffolding types remains unexplored.

\subsection{Future Work}

Several directions for future research emerge from this study's findings and limitations. First, larger-scale studies with more diverse samples---particularly true novices in introductory programming courses---would strengthen the generalizability of findings about scaffolding preferences and effectiveness. Such studies could employ experimental designs with control conditions to establish causal relationships between scaffolding availability and learning outcomes.

Second, the mental model mismatch phenomenon observed in the case study analysis warrants systematic investigation. Future systems could implement detection mechanisms that identify when a learner's approach diverges from the scaffolding's solution and provide targeted metacognitive prompts such as ``The solution shown uses a different approach than the one you tried. Can you identify how they differ?'' Research should examine whether such explicit approach-comparison prompts help learners recognize when to abandon ineffective strategies.

Third, the Scaffolding Learning Questionnaire should be validated through administration to larger samples, factor analysis to confirm its theoretical structure, and correlation with behavioral measures of scaffolding engagement and learning outcomes. Items probing learners' awareness of approach-level differences between their solution and the scaffolding should be developed and tested.

Fourth, future research should explore adaptive scaffolding that responds to learner behavior in real time. The observation that some learners may frame scaffolding as creating a ``hierarchy'' (restructuring thinking) and some may frame it as providing ``hints'' suggests that scaffolding presentation and framing may be as important as scaffolding content. Systems could experiment with different introductions to scaffolding options that emphasize their role as alternative approaches rather than simply answers.

Fifth, the role of time pressure in scaffolding effectiveness deserves systematic study. Comparing scaffolding engagement and effectiveness under per-problem timing versus problem-set timing versus untimed conditions would clarify whether the delayed help-seeking observed in this study reflects time pressure, help-avoidance tendencies, or some combination.

Sixth, future work should investigate additional scaffolding types that support metacognitive processing, such as problem types requiring learners to generate test cases \cite{Pechorina2023-kd, Wrenn2019-hp} and Prompt problems \cite{Denny2024-kx}. Comparing how learners engage with scaffolding that targets different aspects of the programming process (algorithmic design, implementation, testing, debugging) would inform the design of comprehensive scaffolding systems.

Finally, longitudinal studies examining how scaffolding preferences and effectiveness change as learners gain experience would contribute to understanding developmental trajectories in programming skill acquisition. The expertise reversal effect \cite{Sweller2011-ei} observed in prior Parsons problems research \cite{Ericson2022-hp} suggests that optimal scaffolding may differ for learners at different skill levels; tracking individual learners over time would illuminate these dynamics.

\section{Conclusion}
This study explored learners' perceptions of using Faded Parsons and Pseudocode Parsons problems as optional scaffolding for write-code problems within Codespec, a computer programming practice environment designed to offer multiple means of engagement. Nine programmers solved eleven problems, rated their cognitive load using the Paas scale, completed retrospective think-aloud interviews, and responded to the Scaffolding Learning Questionnaire.

The findings reveal that offering Faded and Pseudocode Parsons problems as optional scaffolds can support learners' progress on write-code tasks through multiple mechanisms: enabling comprehension monitoring by serving as externalized checkpoints for comparing solutions, encouraging strategy formation by exposing learners to alternative problem-solving approaches, and helping refine existing programming knowledge by highlighting syntactic patterns and algorithmic structures. Learners strategically differentiated between scaffolding types, using Faded Parsons problems primarily for syntax and structure support while using Pseudocode Parsons problems for high-level algorithmic reasoning. Some learners also perceived Faded Parsons problems as a desirable difficulty---the fill-in-the-blanks feature maintained productive challenge while still providing scaffolded support.

This study contributes to the field of computing education and human-computer interaction in several ways. First, it extends the functionality of problem spaces that support Parsons problems by introducing Pseudocode Parsons problems and demonstrating that different scaffolding types serve complementary functions in learners' problem-solving processes. Second, it provides empirical evidence---albeit preliminary given the sample size---that Faded and Pseudocode Parsons problems can function as effective scaffolding for write-code tasks, building on prior work that established this benefit for traditional Parsons problems~\cite{Hou2022-xa, Hou2023-dg}. Third, it pilots the Scaffolding Learning Questionnaire as an instrument for measuring scaffolding effectiveness, derived from Reiser and Tabak's \cite{Reiser2014-ql} theoretical framework. Fourth, it identifies the mental model mismatch phenomenon as a critical boundary condition for scaffolding effectiveness, with implications for the design of adaptive scaffolding systems.

The practical implications for design are threefold. First, scaffolding systems should provide multiple scaffolding types (e.g., both Faded and Pseudocode Parsons problems) to accommodate learners' varying needs for syntactic versus conceptual support. Second, systems should explicitly surface approach-level differences when a learner's solution diverges from the scaffolding, rather than assuming learners will recognize the discrepancy independently. Third, scaffolding for Faded Parsons problems should include progressive hints for blank completion to prevent learners from becoming stuck at intermediate scaffolding levels while preserving the desirable difficulty that some learners value.

Learning to program is not one-size-fits-all. Learners bring diverse prior knowledge, mental models, and preferences to the task of acquiring programming skills. This study provides evidence that offering multiple scaffolding options---Faded Parsons problems for those seeking syntactic guidance and Pseudocode Parsons problems for those needing algorithmic support---can help meet these diverse needs. Yet it also underscores the complexity of scaffolding. The same support that reduces cognitive load for one learner may increase it for another whose approach diverges from the scaffolding's solution. Effective scaffolding systems must not only provide appropriate support but also help learners recognize when their current approach requires revision---a metacognitive skill that, as this study suggests, remains challenging to scaffold.

\begin{acks}
The author thanks the participants for their time and contributions to this study. The author also acknowledges the use of OpenAI's ChatGPT (i.e., GPT-5.1) and Anthropic's Claude for Education (i.e., Opus 4.5) as AI writing assistants during the preparation of this manuscript.
\end{acks}

\bibliographystyle{ACM-Reference-Format}
\bibliography{paperpile}

@INCOLLECTION{Malmi2019-zu,
  title     = "Tools and Environments",
  author    = "Malmi, Lauri and Utting, Ian and Ko, Amy J",
  editor    = "Fincher, Sally A and Robins, Anthony V",
  booktitle = "\textit{The Cambridge Handbook of Computing Education Research}",
  publisher = "Cambridge University Press",
  address   = "Cambridge, England",
  pages     = "639--662",
  year      =  2019
}

@INCOLLECTION{Robins2019-po,
  title     = "Novice Programmers and Introductory Programming",
  author    = "Robins, Anthony V",
  editor    = "Fincher, Sally A and Robins, Anthony V",
  booktitle = "\textit{The Cambridge Handbook of Computing Education Research}",
  publisher = "Cambridge University Press",
  address   = "Cambridge, England",
  pages     = "327--376",
  year      =  2019
}

@ARTICLE{Vygotsky1978-lj,
  title     = "Mind in society: The development of higher psychological
               processes",
  author    = "Vygotsky, L S",
  publisher = "Harvard university press",
  volume    =  86,
  year      =  1978
}

@INPROCEEDINGS{Wu2024-kx,
  title     = "Evaluating micro parsons problems as exam questions",
  author    = "Wu, Zihan and Smith, IV, David H",
  booktitle = "Proceedings of the 2024 on Innovation and Technology in Computer
               Science Education V. 1",
  publisher = "ACM",
  address   = "New York, NY, USA",
  pages     = "674--680",
  month     =  jul,
  year      =  2024
}

@INPROCEEDINGS{Haynes-Magyar2024-au,
  title     = "Neurodiverse programmers and the accessibility of parsons
               problems: An exploratory multiple-case study",
  author    = "Haynes-Magyar, Carl",
  booktitle = "Proceedings of the 55th ACM Technical Symposium on Computer
               Science Education V. 1",
  publisher = "ACM",
  address   = "New York, NY, USA",
  pages     = "491--497",
  series    = "SIGCSE 2024",
  month     =  mar,
  year      =  2024
}

@INPROCEEDINGS{Denny2024-kx,
  title     = "Prompt problems: A new programming exercise for the generative
               {AI} era",
  author    = "Denny, Paul and Leinonen, Juho and Prather, James and
               Luxton-Reilly, Andrew and Amarouche, Thezyrie and Becker, Brett A
               and Reeves, Brent N",
  booktitle = "Proceedings of the 55th ACM Technical Symposium on Computer
               Science Education V. 1",
  publisher = "ACM",
  address   = "New York, NY, USA",
  pages     = "296--302",
  month     =  mar,
  year      =  2024
}

@INCOLLECTION{Reiser2014-ql,
  title     = "Scaffolding",
  author    = "Reiser, Brian J and Tabak, Iris",
  editor    = "Sawyer, R Keith",
  booktitle = "The Cambridge Handbook of the Learning Sciences",
  publisher = "Harvard University Press",
  address   = "Cambridge, MA",
  edition   = "Second",
  pages     = "44--62",
  year      =  2014
}

@INPROCEEDINGS{Garcia2021-jt,
  title     = "Evaluating parsons problems as a design-based intervention",
  author    = "Garcia, Rita",
  booktitle = "2021 IEEE Frontiers in Education Conference (FIE)",
  publisher = "IEEE",
  pages     = "1--9",
  month     =  oct,
  year      =  2021
}

@ARTICLE{Chen2018-xp,
  title     = "Undesirable difficulty effects in the learning of high-element
               interactivity materials",
  author    = "Chen, Ouhao and Castro-Alonso, Juan C and Paas, Fred and Sweller,
               John",
  journal   = "Front. Psychol.",
  publisher = "Frontiers Media SA",
  volume    =  9,
  pages     =  1483,
  month     =  aug,
  year      =  2018,
  language  = "en"
}

@ARTICLE{Ihantola2011-np,
  title     = "Two-dimensional parson’s puzzles: The concept, tools, and first
               observations",
  author    = "Ihantola, Petri and Karavirta, Ville",
  journal   = "J. Inf. Technol. Educ. Innov. Pract.",
  publisher = "Informing Science Institute",
  volume    =  10,
  pages     = "119--132",
  year      =  2011,
  language  = "en"
}

@INPROCEEDINGS{Smith2023-pv,
  title     = "Investigating the role and impact of distractors on parsons
               problems in {CS1} assessments",
  author    = "Smith, IV, David H and Fowler, Max and Zilles, Craig",
  booktitle = "Proceedings of the 2023 Conference on Innovation and Technology
               in Computer Science Education V. 1",
  publisher = "ACM",
  address   = "New York, NY, USA",
  month     =  jun,
  year      =  2023
}

@ARTICLE{Kelleher2005-ce,
  title     = "Lowering the barriers to programming: A taxonomy of programming
               environments and languages for novice programmers",
  author    = "Kelleher, Caitlin and Pausch, Randy",
  journal   = "ACM Comput. Surv.",
  publisher = "Association for Computing Machinery (ACM)",
  address   = "New York, NY, USA",
  volume    =  37,
  number    =  2,
  pages     = "83--137",
  month     =  jun,
  year      =  2005,
  language  = "en"
}

@INPROCEEDINGS{Luxton-Reilly2016-bm,
  title     = "Learning to program is easy",
  author    = "Luxton-Reilly, Andrew",
  booktitle = "Proceedings of the 2016 ACM Conference on Innovation and
               Technology in Computer Science Education",
  publisher = "ACM",
  address   = "New York, NY, USA",
  pages     = "284--289",
  series    = "ITiCSE '16",
  month     =  jul,
  year      =  2016
}

@INPROCEEDINGS{Smith2023-em,
  title     = "Comparing the impacts of visually grouped and jumbled distractors
               on parsons problems in {CS1} assessments",
  author    = "Smith, IV, David H and Poulsen, Seth and Fowler, Max and Zilles,
               Craig",
  booktitle = "Proceedings of the ACM Conference on Global Computing Education
               Vol 1",
  publisher = "ACM",
  address   = "New York, NY, USA",
  pages     = "154--160",
  series    = "CompEd 2023",
  month     =  dec,
  year      =  2023
}

@ARTICLE{Du-Boulay1986-sl,
  title     = "Some difficulties of learning to program",
  author    = "Du Boulay, Benedict",
  journal   = "J. Educ. Comput. Res.",
  publisher = "SAGE Publications",
  volume    =  2,
  number    =  1,
  pages     = "57--73",
  month     =  feb,
  year      =  1986,
  language  = "en"
}

@INPROCEEDINGS{Ericson2022-ab,
  title     = "Parsons problems and beyond: Systematic literature review and
               empirical study designs",
  author    = "Ericson, Barbara J and Denny, Paul and Prather, James and Duran,
               Rodrigo and Hellas, Arto and Leinonen, Juho and Miller, Craig S
               and Morrison, Briana B and Pearce, Janice L and Rodger, Susan H",
  booktitle = "Proceedings of the 2022 Working Group Reports on Innovation and
               Technology in Computer Science Education",
  publisher = "ACM",
  address   = "New York, NY, USA",
  pages     = "191--234",
  series    = "ITiCSE-WGR '22",
  month     =  dec,
  year      =  2022
}

@INPROCEEDINGS{Harms2016-bx,
  title     = "Distractors in parsons problems decrease learning efficiency for
               young novice programmers",
  author    = "Harms, Kyle James and Chen, Jason and Kelleher, Caitlin L",
  booktitle = "Proceedings of the 2016 ACM Conference on International Computing
               Education Research",
  publisher = "ACM",
  address   = "New York, NY, USA",
  pages     = "241--250",
  series    = "ICER '16",
  month     =  aug,
  year      =  2016
}

@INPROCEEDINGS{Zavaleta-Bernuy2020-ra,
  title     = "What are we asking our students? A literature map of student
               surveys in computer science education",
  author    = "Zavaleta Bernuy, Angela and Harrington, Brian",
  booktitle = "Proceedings of the 2020 ACM Conference on Innovation and
               Technology in Computer Science Education",
  publisher = "ACM",
  address   = "New York, NY, USA",
  pages     = "418--424",
  series    = "ITiCSE '20",
  month     =  jun,
  year      =  2020
}

@ARTICLE{Paas2010-xi,
  title     = "Cognitive load theory: New conceptualizations, specifications,
               and integrated research perspectives",
  author    = "Paas, Fred and van Gog, Tamara and Sweller, John",
  journal   = "Educ. Psychol. Rev.",
  publisher = "Springer Nature",
  volume    =  22,
  number    =  2,
  pages     = "115--121",
  month     =  jun,
  year      =  2010,
  language  = "en"
}

@INPROCEEDINGS{Haynes2021-qd,
  title     = "Problem-solving efficiency and cognitive load for adaptive
               parsons problems vs. Writing the equivalent code",
  author    = "Haynes, Carl C and Ericson, Barbara J",
  booktitle = "Proceedings of the 2021 CHI Conference on Human Factors in
               Computing Systems",
  publisher = "ACM",
  address   = "New York, NY, USA",
  number    = "Article 60",
  pages     = "1--15",
  series    = "CHI '21",
  month     =  may,
  year      =  2021
}

@INPROCEEDINGS{Weinman2021-in,
  title     = "Improving instruction of programming patterns with faded parsons
               problems",
  author    = "Weinman, Nathaniel and Fox, Armando and Hearst, Marti A",
  booktitle = "Proceedings of the 2021 CHI Conference on Human Factors in
               Computing Systems",
  publisher = "ACM",
  address   = "New York, NY, USA",
  number    = "Article 53",
  pages     = "1--4",
  series    = "CHI '21",
  month     =  may,
  year      =  2021
}

@ARTICLE{Lister2010-fi,
  title     = "Naturally occurring data as research instrument: analyzing
               examination responses to study the novice programmer",
  author    = "Lister, Raymond and Clear, Tony and {Simon} and Bouvier, Dennis J
               and Carter, Paul and Eckerdal, Anna and Jacková, Jana and Lopez,
               Mike and McCartney, Robert and Robbins, Phil and Seppälä, Otto
               and Thompson, Errol",
  journal   = "SIGCSE Bull.",
  publisher = "Association for Computing Machinery (ACM)",
  address   = "New York, NY, USA",
  volume    =  41,
  number    =  4,
  pages     = "156--173",
  month     =  jan,
  year      =  2010,
  language  = "en"
}

@INPROCEEDINGS{Ericson2018-bp,
  title     = "Evaluating the efficiency and effectiveness of adaptive parsons
               problems",
  author    = "Ericson, Barbara J and Foley, James D and Rick, Jochen",
  booktitle = "Proceedings of the 2018 ACM Conference on International Computing
               Education Research",
  publisher = "ACM",
  address   = "New York, NY, USA",
  pages     = "60--68",
  series    = "ICER '18",
  month     =  aug,
  year      =  2018
}

@INPROCEEDINGS{Ericson2019-ce,
  title     = "Investigating the affect and effect of adaptive parsons problems",
  author    = "Ericson, Barbara and McCall, Austin and Cunningham, Kathryn",
  booktitle = "Proceedings of the 19th Koli Calling International Conference on
               Computing Education Research",
  publisher = "ACM",
  address   = "New York, NY, USA",
  number    = "Article 6",
  pages     = "1--10",
  series    = "Koli Calling '19",
  month     =  nov,
  year      =  2019
}

@ARTICLE{Parsons2006-tw,
  title     = "Parson's programming puzzles: a fun and effective learning tool
               for first programming courses",
  author    = "Parsons, D and Haden, P",
  journal   = "of the 8th Australasian Conference on …",
  publisher = "researchgate.net",
  pages     = "157--163",
  year      =  2006
}

@INPROCEEDINGS{Holden2003-oi,
  title     = "The impact of prior experience in an information technology
               programming course sequence",
  author    = "Holden, Edward and Weeden, Elissa",
  booktitle = "Proceedings of the 4th conference on Information technology
               curriculum",
  publisher = "ACM",
  address   = "New York, NY, USA",
  pages     = "41--46",
  series    = "CITC4 '03",
  month     =  oct,
  year      =  2003
}

@ARTICLE{Siegmund2014-zj,
  title     = "Measuring and modeling programming experience",
  author    = "Siegmund, Janet and Kästner, Christian and Liebig, Jörg and Apel,
               Sven and Hanenberg, Stefan",
  journal   = "Empir. Softw. Eng.",
  publisher = "Springer Science and Business Media LLC",
  volume    =  19,
  number    =  5,
  pages     = "1299--1334",
  month     =  oct,
  year      =  2014,
  language  = "en"
}

@INPROCEEDINGS{Loksa2020-cy,
  title     = "Investigating novices' in situ reflections on their programming
               process",
  author    = "Loksa, Dastyni and Xie, Benjamin and Kwik, Harrison and Ko, Amy J",
  booktitle = "Proceedings of the 51st ACM Technical Symposium on Computer
               Science Education",
  publisher = "ACM",
  address   = "New York, NY, USA",
  pages     = "149--155",
  series    = "SIGCSE '20",
  month     =  feb,
  year      =  2020
}

@INPROCEEDINGS{Wrenn2019-hp,
  title     = "Executable examples for programming problem comprehension",
  author    = "Wrenn, John and Krishnamurthi, Shriram",
  booktitle = "Proceedings of the 2019 ACM Conference on International Computing
               Education Research",
  publisher = "ACM",
  address   = "New York, NY, USA",
  pages     = "131--139",
  series    = "ICER '19",
  month     =  jul,
  year      =  2019
}

@INPROCEEDINGS{Du2020-ic,
  title     = "A review of research on parsons problems",
  author    = "Du, Yuemeng and Luxton-Reilly, Andrew and Denny, Paul",
  booktitle = "Proceedings of the Twenty-Second Australasian Computing Education
               Conference",
  publisher = "ACM",
  address   = "New York, NY, USA",
  pages     = "195--202",
  series    = "ACE'20",
  month     =  feb,
  year      =  2020
}

@ARTICLE{Lin2021-dx,
  title     = "The landscape of Block-based programming: Characteristics of
               block-based environments and how they support the transition to
               text-based programming",
  author    = "Lin, Yuhan and Weintrop, David",
  journal   = "J. Comput. Lang.",
  publisher = "Elsevier BV",
  volume    =  67,
  number    =  101075,
  pages     =  101075,
  month     =  dec,
  year      =  2021,
  language  = "en"
}

@INPROCEEDINGS{Helminen2013-su,
  title     = "How do students solve parsons programming problems? --
               execution-based vs. Line-based feedback",
  author    = "Helminen, J and Ihantola, P and Karavirta, V and Alaoutinen, S",
  booktitle = "2013 Learning and Teaching in Computing and Engineering",
  publisher = "IEEE",
  pages     = "55--61",
  month     =  mar,
  year      =  2013
}

@INPROCEEDINGS{Haynes-Magyar2022-tr,
  title     = "The impact of solving adaptive parsons problems with common and
               uncommon solutions",
  author    = "Haynes-Magyar, Carl and Ericson, Barbara",
  booktitle = "Proceedings of the 22nd Koli Calling International Conference on
               Computing Education Research",
  publisher = "ACM",
  address   = "New York, NY, USA",
  number    = "Article 23",
  pages     = "1--14",
  series    = "Koli Calling '22",
  month     =  nov,
  year      =  2022
}

@ARTICLE{Sweller2019-ag,
  title     = "Cognitive architecture and instructional design: 20 years later",
  author    = "Sweller, John and van Merriënboer, Jeroen J G and Paas, Fred",
  journal   = "Educ. Psychol. Rev.",
  publisher = "Springer Science and Business Media LLC",
  volume    =  31,
  number    =  2,
  pages     = "261--292",
  month     =  jun,
  year      =  2019,
  language  = "en"
}

@INPROCEEDINGS{Denny2008-re,
  title     = "Evaluating a new exam question: Parsons problems",
  author    = "Denny, Paul and Luxton-Reilly, Andrew and Simon, Beth",
  booktitle = "Proceedings of the Fourth international Workshop on Computing
               Education Research",
  publisher = "ACM",
  address   = "New York, NY, USA",
  pages     = "113--124",
  series    = "ICER '08",
  month     =  sep,
  year      =  2008
}

@INPROCEEDINGS{Morrison2014-zv,
  title     = "Measuring cognitive load in introductory {CS}: adaptation of an
               instrument",
  author    = "Morrison, Briana B and Dorn, Brian and Guzdial, Mark",
  booktitle = "Proceedings of the tenth annual conference on International
               computing education research",
  publisher = "ACM",
  address   = "New York, NY, USA",
  pages     = "131--138",
  series    = "ICER '14",
  month     =  jul,
  year      =  2014
}

@INPROCEEDINGS{Zavgorodniaia2020-uw,
  title     = "Measuring the cognitive load of learning to program: A
               replication study",
  author    = "Zavgorodniaia, Albina and Duran, Rodrigo and Hellas, Arto and
               Seppala, Otto and Sorva, Juha",
  booktitle = "United Kingdom \& Ireland Computing Education Research conference",
  publisher = "ACM",
  address   = "New York, NY, USA",
  pages     = "3--9",
  series    = "UKICER '20",
  month     =  sep,
  year      =  2020
}

@INPROCEEDINGS{Barr2023-pr,
  title     = "{CS+} {X}: Approaches, Challenges, and Opportunities in
               Developing Interdisciplinary Computing Curricula",
  author    = "Barr, Valerie and Brodley, Carla E and Gunter, Elsa L and
               Guzdial, Mark and Libeskind-Hadas, Ran and Manaris, Bill",
  booktitle = "Proceedings of ACM CS202X",
  publisher = "csed.acm.org",
  year      =  2023
}

@INPROCEEDINGS{Hou2023-dg,
  title     = "Understanding the effects of using parsons problems to scaffold
               code writing for students with varying {CS} self-efficacy levels",
  author    = "Hou, Xinying and Ericson, Barbara Jane and Wang, Xu",
  booktitle = "Proceedings of the 23rd Koli Calling International Conference on
               Computing Education Research",
  publisher = "ACM",
  address   = "New York, NY, USA",
  volume    =  1,
  number    = "Article 1",
  pages     = "1--12",
  series    = "Koli Calling '23",
  month     =  nov,
  year      =  2023
}

@INPROCEEDINGS{Pechorina2023-kd,
  title     = "Metacodenition: Scaffolding the problem-solving process for
               novice programmers",
  author    = "Pechorina, Yulia and Anderson, Keith and Denny, Paul",
  booktitle = "Proceedings of the 25th Australasian Computing Education
               Conference",
  publisher = "ACM",
  address   = "New York, NY, USA",
  pages     = "59--68",
  series    = "ACE '23",
  month     =  jan,
  year      =  2023
}

@INPROCEEDINGS{Fromont2023-rn,
  title     = "Exploring the difficulty of faded parsons problems for
               programming education",
  author    = "Fromont, Flynn and Jayamanne, Hiruna and Denny, Paul",
  booktitle = "Australasian Computing Education Conference",
  publisher = "ACM",
  address   = "New York, NY, USA",
  pages     = "113--122",
  series    = "ACE '23",
  month     =  jan,
  year      =  2023
}

@ARTICLE{Paas1992-mk,
  title     = "Training strategies for attaining transfer of problem-solving
               skill in statistics: A cognitive-load approach",
  author    = "Paas, Fred G",
  journal   = "J. Educ. Psychol.",
  publisher = "American Psychological Association (APA)",
  volume    =  84,
  number    =  4,
  pages     = "429--434",
  month     =  dec,
  year      =  1992
}

@INPROCEEDINGS{Garcia2018-at,
  title     = "Scaffolding the Design Process using Parsons Problems",
  author    = "Garcia, Rita and Falkner, Katrina and Vivian, Rebecca",
  booktitle = "Proceedings of the 18th Koli Calling International Conference on
               Computing Education Research",
  publisher = "ACM",
  address   = "New York, NY, USA",
  number    = "Article 26",
  pages     = "1--2",
  series    = "Koli Calling '18",
  month     =  nov,
  year      =  2018
}

@INPROCEEDINGS{Hou2022-xa,
  title     = "Using adaptive parsons problems to scaffold write-code problems",
  author    = "Hou, Xinying and Ericson, Barbara Jane and Wang, Xu",
  booktitle = "Proceedings of the 2022 ACM Conference on International Computing
               Education Research - Volume 1",
  publisher = "ACM",
  address   = "New York, NY, USA",
  volume    =  1,
  pages     = "15--26",
  series    = "ICER '22",
  month     =  aug,
  year      =  2022
}

@INPROCEEDINGS{Wu2023-ui,
  title     = "Using micro parsons problems to scaffold the learning of regular
               expressions",
  author    = "Wu, Zihan and Ericson, Barbara J and Brooks, Christopher",
  booktitle = "Proceedings of the 2023 Conference on Innovation and Technology
               in Computer Science Education V. 1",
  publisher = "ACM",
  address   = "New York, NY, USA",
  pages     = "457--463",
  series    = "ITiCSE 2023",
  month     =  jun,
  year      =  2023
}

@INPROCEEDINGS{Smith2023-ep,
  title     = "Discovering, autogenerating, and evaluating distractors for
               python parsons problems in {CS1}",
  author    = "Smith, IV, David H and Zilles, Craig",
  booktitle = "Proceedings of the 54th ACM Technical Symposium on Computer
               Science Education V. 1",
  publisher = "ACM",
  address   = "New York, NY, USA",
  volume    =  1,
  pages     = "924--930",
  series    = "SIGCSE 2023",
  month     =  mar,
  year      =  2023
}

@INPROCEEDINGS{Ericson2022-hp,
  title     = "Adaptive parsons problems as active learning activities during
               lecture",
  author    = "Ericson, Barbara and Haynes-Magyar, Carl",
  booktitle = "Proceedings of the 27th ACM Conference on on Innovation and
               Technology in Computer Science Education Vol. 1",
  publisher = "ACM",
  address   = "New York, NY, USA",
  pages     = "290--296",
  series    = "ITiCSE '22",
  month     =  jul,
  year      =  2022
}

@INPROCEEDINGS{Morrison2016-qd,
  title     = "Subgoals help students solve parsons problems",
  author    = "Morrison, Briana B and Margulieux, Lauren E and Ericson, Barbara
               and Guzdial, Mark",
  booktitle = "Proceedings of the 47th ACM Technical Symposium on Computing
               Science Education",
  publisher = "ACM",
  address   = "New York, NY, USA",
  pages     = "42--47",
  series    = "SIGCSE '16",
  month     =  feb,
  year      =  2016
}

@ARTICLE{Xie2019-nr,
  title     = "A theory of instruction for introductory programming skills",
  author    = "Xie, Benjamin and Loksa, Dastyni and Nelson, Greg L and Davidson,
               Matthew J and Dong, Dongsheng and Kwik, Harrison and Tan, Alex
               Hui and Hwa, Leanne and Li, Min and Ko, Amy J",
  journal   = "Comput. Sci. Educ.",
  publisher = "Informa UK Limited",
  volume    =  29,
  number    = "2-3",
  pages     = "205--253",
  month     =  jul,
  year      =  2019,
  language  = "en"
}

@INPROCEEDINGS{Ericson2017-zo,
  title     = "Solving parsons problems versus fixing and writing code",
  author    = "Ericson, Barbara J and Margulieux, Lauren E and Rick, Jochen",
  booktitle = "Proceedings of the 17th Koli Calling International Conference on
               Computing Education Research",
  publisher = "ACM",
  address   = "New York, NY, USA",
  pages     = "20--29",
  series    = "Koli Calling '17",
  month     =  nov,
  year      =  2017
}

@ARTICLE{Bellamy1994-wo,
  title     = "What does pseudo-code do? A psychological analysis of the use of
               pseudo-code by experienced programmers",
  author    = "Bellamy, Rachel",
  journal   = "Hum.-Comput. Interact.",
  publisher = "Informa UK Limited",
  volume    =  9,
  number    =  2,
  pages     = "225--246",
  month     =  jun,
  year      =  1994,
  language  = "en"
}

@INPROCEEDINGS{Wu2024-ni,
  title     = "{SQL} puzzles: Evaluating micro parsons problems with different
               feedbacks as practice for novices",
  author    = "Wu, Zihan and Ericson, Barbara J",
  booktitle = "Proceedings of the CHI Conference on Human Factors in Computing
               Systems",
  publisher = "ACM",
  address   = "New York, NY, USA",
  number    = "Article 930",
  pages     = "1--15",
  series    = "CHI '24",
  month     =  may,
  year      =  2024
}

@ARTICLE{Aleven2002-im,
  title     = "An effective metacognitive strategy: learning by doing and
               explaining with a computer‐based Cognitive Tutor",
  author    = "Aleven, Vincent A W M M and Koedinger, Kenneth R",
  journal   = "Cogn. Sci.",
  publisher = "Wiley",
  volume    =  26,
  number    =  2,
  pages     = "147--179",
  month     =  mar,
  year      =  2002,
  language  = "en"
}

@INPROCEEDINGS{Allen2019-uj,
  title     = "An analysis of using many small programs in {CS1}",
  author    = "Allen, Joe Michael and Vahid, Frank and Edgcomb, Alex and Downey,
               Kelly and Miller, Kris",
  booktitle = "Proceedings of the 50th ACM Technical Symposium on Computer
               Science Education",
  publisher = "ACM",
  address   = "New York, NY, USA",
  pages     = "585--591",
  series    = "SIGCSE '19",
  month     =  feb,
  year      =  2019
}

@INCOLLECTION{Metcalf2012-ey,
  title     = "{MODEL}-{IT} : A Design Retrospective",
  author    = "Metcalf, S J and Krajcik, J and Soloway, E",
  editor    = "Jacobson, Michael J and Kozma, Robert B",
  booktitle = "Innovations in Science and Mathematics Education",
  publisher = "Routledge",
  pages     = "89--127",
  month     =  dec,
  year      =  2012
}

@INCOLLECTION{Hart1988-tc,
  title     = "Development of {NASA}-{TLX} (task load index): Results of
               empirical and theoretical research",
  author    = "Hart, Sandra G and Staveland, Lowell E",
  booktitle = "Advances in Psychology",
  publisher = "Elsevier",
  pages     = "139--183",
  series    = "Advances in Psychology",
  year      =  1988,
  language  = "en"
}

@ARTICLE{Eysink2009-kl,
  title     = "Learner performance in multimedia learning arrangements: An
               analysis across instructional approaches",
  author    = "Eysink, Tessa H S and de Jong, Ton and Berthold, Kirsten and
               Kolloffel, Bas and Opfermann, Maria and Wouters, Pieter",
  journal   = "Am. Educ. Res. J.",
  publisher = "American Educational Research Association (AERA)",
  volume    =  46,
  number    =  4,
  pages     = "1107--1149",
  month     =  dec,
  year      =  2009,
  language  = "en"
}

@ARTICLE{Swaak2001-rf,
  title     = "Learner vs. System control in using online support for
               simulation-based discovery learning",
  author    = "Swaak, Janine and de Jong, Ton",
  journal   = "Learn. Environ. Res.",
  publisher = "Springer Science and Business Media LLC",
  volume    =  4,
  number    =  3,
  pages     = "217--241",
  month     =  oct,
  year      =  2001,
  language  = "en"
}

@INCOLLECTION{Sweller2011-ei,
  title     = "The expertise reversal effect",
  author    = "Sweller, John and Ayres, Paul and Kalyuga, Slava",
  booktitle = "Cognitive Load Theory",
  publisher = "Springer New York",
  address   = "New York, NY",
  pages     = "155--170",
  year      =  2011
}

@ARTICLE{Oakeson2025-ev,
  title         = "Choose your own solution: Supporting optional blocks in block
                   ordering problems",
  author        = "Oakeson, Skyler and Smith, IV, David H and Winder, Jaxton and
                   Poulsen, Seth",
  journal       = "arXiv [cs.HC]",
  month         =  oct,
  year          =  2025,
  archivePrefix = "arXiv",
  primaryClass  = "cs.HC"
}

@INPROCEEDINGS{Smith2024-cz,
  title     = "Distractors make you pay attention: Investigating the learning
               outcomes of including distractor blocks in parsons problems",
  author    = "Smith, David H and Poulsen, Seth and Emeka, Chinedu and Wu, Zihan
               and Haynes-Magyar, Carl and Zilles, Craig",
  booktitle = "Proceedings of the 2024 ACM Conference on International Computing
               Education Research - Volume 1",
  publisher = "ACM",
  address   = "New York, NY, USA",
  volume    =  10,
  pages     = "177--191",
  month     =  aug,
  year      =  2024
}

@ARTICLE{Catrambone1998-ka,
  title     = "The subgoal learning model: Creating better examples so that
               students can solve novel problems",
  author    = "Catrambone, Richard",
  journal   = "J. Exp. Psychol. Gen.",
  publisher = "American Psychological Association (APA)",
  volume    =  127,
  number    =  4,
  pages     = "355--376",
  month     =  dec,
  year      =  1998,
  language  = "en"
}

@INPROCEEDINGS{Murphy2012-ey,
  title     = "Ability to 'explain in plain english' linked to proficiency in
               computer-based programming",
  author    = "Murphy, Laurie and Fitzgerald, Sue and Lister, Raymond and
               McCauley, Renée",
  booktitle = "Proceedings of the ninth annual international conference on
               International computing education research",
  publisher = "ACM",
  address   = "New York, NY, USA",
  month     =  sep,
  year      =  2012
}

@INPROCEEDINGS{Smith2024-tr,
  title     = "Prompting for comprehension: Exploring the intersection of
               explain in plain English questions and prompt writing",
  author    = "Smith, IV, David H and Denny, Paul and Fowler, Max",
  booktitle = "Proceedings of the Eleventh ACM Conference on Learning @ Scale",
  publisher = "ACM",
  address   = "New York, NY, USA",
  pages     = "39--50",
  month     =  jul,
  year      =  2024
}

@ARTICLE{van-Merrienboer2005-eb,
  title     = "Cognitive load theory and complex learning: Recent developments
               and future directions",
  author    = "van Merriënboer, Jeroen J G and Sweller, John",
  journal   = "Educ. Psychol. Rev.",
  publisher = "Springer Science and Business Media LLC",
  volume    =  17,
  number    =  2,
  pages     = "147--177",
  month     =  jun,
  year      =  2005,
  language  = "en"
}

@INPROCEEDINGS{Hassany2025-kn,
  title     = "Generating effective distractors for introductory programming
               challenges: {LLMs} vs humans",
  author    = "Hassany, Mohammad and Brusilovsky, Peter and Savelka, Jaromir and
               Lekshmi Narayanan, Arun Balajiee and Akhuseyinoglu, Kamil and
               Agarwal, Arav and Hendrawan, Rully Agus",
  booktitle = "Proceedings of the 15th International Learning Analytics and
               Knowledge Conference",
  publisher = "ACM",
  address   = "New York, NY, USA",
  pages     = "484--493",
  month     =  mar,
  year      =  2025
}

@INPROCEEDINGS{Szabo2025-bg,
  title     = "Parsons problems and computing education learning theories",
  author    = "Szabo, Claudia and Sheard, Judy and Malmi, Lauri and Kinnunen,
               Paivi",
  booktitle = "Proceedings of the 25th Koli Calling International Conference on
               Computing Education Research",
  publisher = "ACM",
  address   = "New York, NY, USA",
  pages     = "1--10",
  month     =  nov,
  year      =  2025
}

@ARTICLE{Wood1976-sr,
  title   = "The role of tutoring in problem solving",
  author  = "Wood, D and Bruner, J S and Ross, G",
  journal = "Journal of child psychology and psychiatry",
  volume  =  17,
  number  =  2,
  pages   = "89--100",
  year    =  1976
}

@INPROCEEDINGS{Pradhan-Newar2025-qh,
  title     = "Mining hierarchies with conviction: Constructing the {CS1} skill
               hierarchy with pairwise comparisons over skill distributions",
  author    = "Pradhan Newar, Dip Kiran and Fowler, Max and Smith, IV, David H
               and Poulsen, Seth",
  booktitle = "Proceedings of the 56th ACM Technical Symposium on Computer
               Science Education V. 2",
  publisher = "ACM",
  address   = "New York, NY, USA",
  pages     = "1583--1584",
  month     =  feb,
  year      =  2025
}

@INCOLLECTION{Hannafin2013-wa,
  title     = "Open learning environments: Foundations, methods, and models",
  author    = "Hannafin, Michael and Land, Susan and Oliver, Kevin",
  booktitle = "Instructional-design Theories and Models",
  publisher = "Routledge",
  address   = "New York",
  pages     = "115--140",
  month     =  may,
  year      =  2013,
  language  = "en"
}

@ARTICLE{Hadwin2001-oe,
  title     = "{CoNoteS2}: A software tool for promoting self-regulation",
  author    = "Hadwin, Allyson Fiona and Winne, Philip H",
  journal   = "Educ. Res. Eval.",
  publisher = "Informa UK Limited",
  volume    =  7,
  number    = "2-3",
  pages     = "313--334",
  month     =  sep,
  year      =  2001
}

@ARTICLE{Yeung2000-bs,
  title     = "Toward a Subjective Mental Workload Measure",
  author    = "Yeung, A S and Lee, C F K and Pena, I M and Ryde, J",
  publisher = "ERIC",
  year      =  2000
}

@ARTICLE{Sweller2025-bo,
  title     = "An integrated human cognitive architecture",
  author    = "Sweller, John",
  journal   = "Educ. Psychol. Rev.",
  publisher = "Springer Science and Business Media LLC",
  volume    =  37,
  number    =  4,
  month     =  dec,
  year      =  2025,
  language  = "en"
}

@ARTICLE{Chen2023-gj,
  title   = "A Cognitive Load Theory Approach to Defining and Measuring Task
             Complexity Through Element Interactivity",
  author  = "Chen, Ouhao and Paas, Fred and Sweller, John",
  journal = "Educ Psychol Rev",
  volume  =  35,
  number  =  2,
  month   =  jun,
  year    =  2023
}

@ARTICLE{Sweller2010-uq,
  title     = "Element interactivity and intrinsic, extraneous, and germane
               cognitive load",
  author    = "Sweller, John",
  journal   = "Educ. Psychol. Rev.",
  publisher = "Springer Science and Business Media LLC",
  volume    =  22,
  number    =  2,
  pages     = "123--138",
  month     =  jun,
  year      =  2010,
  language  = "en"
}

@ARTICLE{Chen2015-wr,
  title     = "The worked example effect, the generation effect, and element
               interactivity",
  author    = "Chen, Ouhao and Kalyuga, Slava and Sweller, John",
  journal   = "J. Educ. Psychol.",
  publisher = "American Psychological Association (APA)",
  volume    =  107,
  number    =  3,
  pages     = "689--704",
  month     =  aug,
  year      =  2015,
  language  = "en"
}

@ARTICLE{Klepsch2017-bx,
  title    = "Development and validation of two instruments measuring intrinsic,
              extraneous, and germane cognitive load",
  author   = "Klepsch, Melina and Schmitz, Florian and Seufert, Tina",
  journal  = "Front. Psychol.",
  volume   =  8,
  pages    =  1997,
  month    =  nov,
  year     =  2017,
  language = "en"
}

@INCOLLECTION{Robins2019-iv,
  title     = "Cognitive Sciences for Computing Education",
  author    = "Robins, Anthony V and Margulieux, Lauren E and Morrison, Briana B",
  booktitle = "The Cambridge Handbook of Computing Education Research",
  publisher = "Cambridge University Press",
  pages     = "231--275",
  month     =  feb,
  year      =  2019
}

@ARTICLE{Klepsch2020-er,
  title     = "Understanding instructional design effects by differentiated
               measurement of intrinsic, extraneous, and germane cognitive load",
  author    = "Klepsch, Melina and Seufert, Tina",
  journal   = "Instr. Sci.",
  publisher = "Springer Science and Business Media LLC",
  volume    =  48,
  number    =  1,
  pages     = "45--77",
  month     =  feb,
  year      =  2020,
  language  = "en"
}

@BOOK{Zheng2017-lr,
  title     = "Cognitive load measurement and application: A theoretical
               framework for meaningful research and practice",
  author    = "Zheng, Robert Z",
  editor    = "Zheng, Robert Z",
  publisher = "Routledge",
  address   = "London, England",
  edition   =  1,
  month     =  nov,
  year      =  2017,
  language  = "en"
}

@ARTICLE{Saye2002-lm,
  title     = "Scaffolding critical reasoning about history and social issues in
               multimedia-supported learning environments",
  author    = "Saye, John W and Brush, Thomas",
  journal   = "Educ. Technol. Res. Dev.",
  publisher = "Springer Science and Business Media LLC",
  volume    =  50,
  number    =  3,
  pages     = "77--96",
  month     =  sep,
  year      =  2002,
  language  = "en"
}

@ARTICLE{Azevedo2005-ha,
  title     = "Adaptive human scaffolding facilitates adolescents’
               self-regulated learning with hypermedia",
  author    = "Azevedo, Roger and Cromley, Jennifer G and Winters, Fielding I
               and Moos, Daniel C and Greene, Jeffrey A",
  journal   = "Instr. Sci.",
  publisher = "Springer Science and Business Media LLC",
  volume    =  33,
  number    = "5-6",
  pages     = "381--412",
  month     =  nov,
  year      =  2005,
  language  = "en"
}

@ARTICLE{Kim2011-ms,
  title     = "Scaffolding problem solving in technology-enhanced learning
               environments ({TELEs}): Bridging research and theory with
               practice",
  author    = "Kim, Minchi C and Hannafin, Michael J",
  journal   = "Comput. Educ.",
  publisher = "Elsevier BV",
  volume    =  56,
  number    =  2,
  pages     = "403--417",
  month     =  feb,
  year      =  2011,
  language  = "en"
}

@ARTICLE{Chi2004-vf,
  title     = "Can tutors monitor students' understanding accurately?",
  author    = "Chi, Michelene T H and Siler, Stephanie A and Jeong, Heisawn",
  journal   = "Cogn. Instr.",
  publisher = "Informa UK Limited",
  volume    =  22,
  number    =  3,
  pages     = "363--387",
  month     =  sep,
  year      =  2004
}

@ARTICLE{Chi2001-ix,
  title     = "Learning from human tutoring",
  author    = "Chi, Michelene T H and Siler, Stephanie A and Jeong, Heisawn and
               Yamauchi, Takashi and Hausmann, Robert G",
  journal   = "Cogn. Sci.",
  publisher = "Wiley",
  volume    =  25,
  number    =  4,
  pages     = "471--533",
  month     =  jul,
  year      =  2001,
  language  = "en"
}

@ARTICLE{Huang2025-sj,
  title     = "Promoting elementary school students’ programming learning:
               Effects of metacognitive vs. cognitive scaffolding",
  author    = "Huang, Yi-Pin and Kim, Hoisoo and Pan, Yingying and Zheng,
               Xiao-Li and Tu, Yun-Fang",
  journal   = "J. Res. Technol. Educ.",
  publisher = "Informa UK Limited",
  volume    =  57,
  number    =  4,
  pages     = "914--929",
  month     =  jul,
  year      =  2025,
  language  = "en"
}

@ARTICLE{Bunde2023-rw,
  title     = "{CONVERSATIONS}: Conversation with a prominent propagator: Carl
               Haynes-Magyar",
  author    = "Bunde, David P and Butler, Zack and Hovey, Christopher L and
               Taylor, Cynthia",
  journal   = "ACM Inroads",
  publisher = "Association for Computing Machinery (ACM)",
  volume    =  14,
  number    =  4,
  pages     = "12--16",
  month     =  dec,
  year      =  2023,
  language  = "en"
}

@INPROCEEDINGS{Haynes-Magyar2022-oj,
  title     = "Codespec: A computer programming practice environment",
  author    = "Haynes-Magyar, Carl Christopher and Haynes-Magyar, Nathaniel
               James",
  booktitle = "Proceedings of the 2022 ACM Conference on International Computing
               Education Research - Volume 2",
  publisher = "ACM",
  address   = "New York, NY, USA",
  pages     = "32--34",
  month     =  aug,
  year      =  2022
}

@ARTICLE{Morehouse2025-nm,
  title     = "Responsible data sharing: Identifying and remedying possible
               re-identification of human participants",
  author    = "Morehouse, Kirsten N and Kurdi, Benedek and Nosek, Brian A",
  journal   = "Am. Psychol.",
  publisher = "American Psychological Association (APA)",
  volume    =  80,
  number    =  6,
  pages     = "928--941",
  month     =  sep,
  year      =  2025,
  language  = "en"
}

@ARTICLE{Phillips2016-ka,
  title     = "Thinking styles and decision making: A meta-analysis",
  author    = "Phillips, Wendy J and Fletcher, Jennifer M and Marks, Anthony D G
               and Hine, Donald W",
  journal   = "Psychol. Bull.",
  publisher = "American Psychological Association (APA)",
  volume    =  142,
  number    =  3,
  pages     = "260--290",
  month     =  mar,
  year      =  2016,
  language  = "en"
}

\appendix





\end{document}